%% file: ms.tex
\documentclass{ar-1col}

\usepackage[comma]{natbib}
\usepackage{color,comment}
\usepackage{amsmath}

\newcommand{\mnras}{\textit{MNRAS }}
\newcommand{\aj}{\textit{Astron.\ J.\ }}
\newcommand{\apj}{\textit{Ap.\ J.\ }}
\newcommand{\apjl}{\textit{Ap.\ J.\ Lett.\ }}
\newcommand{\apjs}{\textit{Ap.\ J.\ Suppl.\ }}
\newcommand{\araa}{\textit{Annu.\ Rev.\ Astron.\ Astrophys.\ }}
\newcommand{\nat}{\textit{Nature }}
\newcommand{\pasp}{\textit {Publ.\ Astron.\ Soc.\ Pac.\ }}
\newcommand{\pasj}{\textit {Publ.\ Astron.\ Soc.\ Jpn.\ }}
\newcommand{\aap}{\textit{Astron.\ Astrophys.\ }}
\jname{Xxxx. Xxx. Xxx. Xxx.}
\jvol{AA}
\jyear{YYYY}
\doi{10.1146/((please add article doi))}

\begin{document}

\markboth{Salim \& Narayanan}{Dust Attenuation in Galaxies}

\title{The Dust Attenuation Law in Galaxies}

\author{Samir Salim$^1$ \& Desika Narayanan$^{2,3}$ 
\affil{$^1$Department of Astronomy, Indiana University, Bloomington, IN, 47404\\ email: salims@indiana.edu}
\affil{$^2$Department of Astronomy, University of Florida, Gainesville, FL,  32611\\ email: desika.narayanan@ufl.edu}
\affil{$^3$Cosmic Dawn Centre at the Niels Bohr Institute, University of Copenhagen \\ and DTU-Space, Technical University of Denmark}}

\begin{abstract}
Understanding the properties and physical mechanisms that shape dust attenuation curves in galaxies is one of the fundamental questions of extragalactic astrophysics, with a great practical significance for deriving the physical properties of galaxies, such as the star formation rate and stellar mass. Attenuation curves result from a combination of dust grain properties, dust content, and the spatial arrangement of dust and different populations of stars.  In this review we assess the current state of the field, paying particular attention to the importance of extinction curves as the building blocks of attenuation laws.   We introduce a quantitative framework to characterize and compare extinction and attenuation curves, present a theoretical foundation for interpreting empirical results, overview an array of observational methods, and review the observational state of the field at both low and high redshift.  Our main conclusions are: Attenuation curves exhibit a large range of slopes, from curves with shallow (Milky Way-like) slopes to those exceeding the slope of the SMC extinction curve.
The slopes of the curves correlate strongly with the effective optical opacities, in the sense that galaxies with low dust column density (lower visual attenuation) tend to have steeper slopes, whereas the galaxies with high dust column density have shallower (grey) slopes. Galaxies appear to exhibit a diverse range of $2175$ \AA\ UV bump strengths, but on average have suppressed bumps compared to the average Milky Way extinction curve. Theoretical studies indicate that variations in bump strength may result from similar geometric and radiative transfer effects that drive the correlation between the slope and the dust column. 
\end{abstract}

\maketitle

\tableofcontents

\section{\MakeUppercase{Introduction}}

Astrophysical dust is one of the key components of the interstellar medium (ISM) of galaxies. Dust particles (hereafter ``grains") originate as a product of stellar evolution. Grains form in the atmospheres of evolved stars or remnants of supernovae, and are then released or ejected into the ISM \citep{Draine2011}, where their masses and sizes evolve owing to a diverse set of physical processes \citep{Asano2013}.  Dust grains typically range in size from 5 to 250 nm \citep{Weingartner2001}, which is two orders of magnitude smaller than household dust particles and comparable to combustion (smoke) particles. The two basic compositions of interstellar dust grains are carbonaceous, including graphite and polycyclic aromatic hydrocarbons (PAHs), and silicate. 

Observationally,  the presence of dust in galaxies is revealed by two general effects: (1) dust produces emission in the infrared (IR) part of the spectrum, consisting of the continuum and various emission and absorption features, and  (2) it modifies the light from the stellar continuum at all wavelengths, in particular the ultraviolet (UV) to near-infrared (NIR) flux from stars. This review focuses on understanding the driving physics and observational properties of the latter effect, encapsulated into two terms that we will define in \S~\ref{section:extinction_attenuation_def}: ``extinction" and ``attenuation."  Throughout, we utilize the terminology ``curves" or ``laws" interchangeably to refer to the wavelength dependence of the optical depth in either extinction or attenuation.    We gear this review towards general practitioners of extragalactic and galaxy evolution studies who wish to use attenuation laws to further our understanding of the physics of dust grains itself, as well as to researchers interested in using attenuation laws from a purely phenomenological perspective in order to correct the observed UV-NIR stellar emission for attenuation.

Placing this review into context, we note that \citet{Savage1979, Mathis1990} and \citet{Draine2003}  reviewed the properties of extinction curves in galaxies, while a foundational review of dust attenuation was presented in \citet{Calzetti2001}. Here, we cover both theoretical and observational progress since these reviews, paying special attention to the details of methodology, which may be critical for resolving some of the current conflicting results. Other aspects of dust in the ISM, including its distribution, evolution, and signatures across the electromagnetic spectrum have been reviewed recently by \citet{Galliano2018}. \citet{Steinacker2013} provide a review of dust radiative transfer methods required to model observational data.  Finally, \citet{Walcher2011} and \citet{Conroy2013} review the practice of fitting galaxy SEDs, in which dust attenuation plays a fundamental role.

\begin{figure}
    \centering
    \includegraphics[scale=0.35]{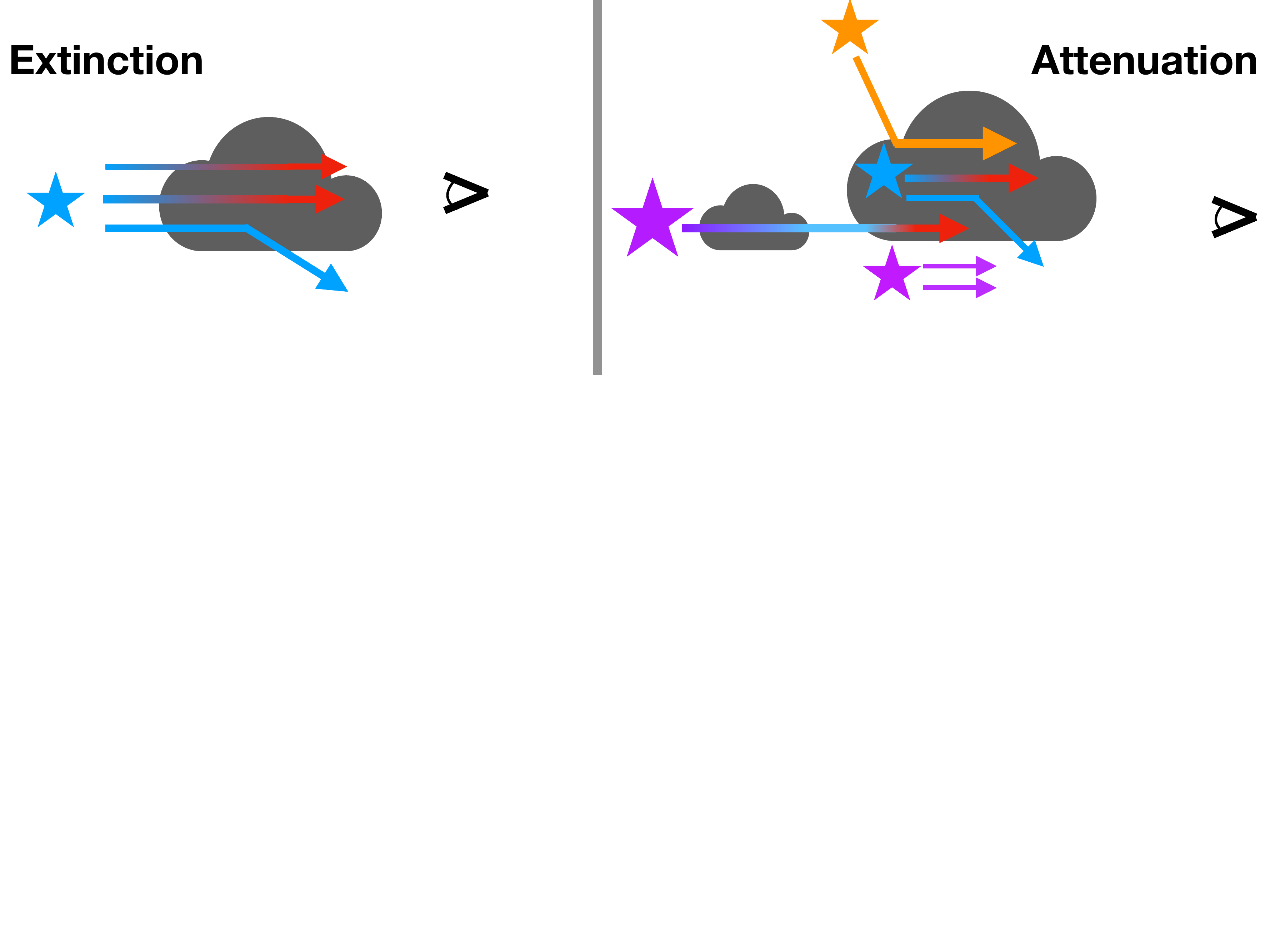}
    \vspace{-6.5cm}
    \caption{Schematic summarizing the difference between extinction and attenuation.  The former encapsulates absorption and scattering out of the line of sight, while the latter folds in the complexities of star-dust geometry in galaxies, and may include scattering back into the line of sight, varying column densities/optical depths, and the contribution by unobscured stars.}
    \label{fig:attenuation_extinction_schematic}
\end{figure}

\subsection{Extinction vs.\ Attenuation}
\label{section:extinction_attenuation_def}
\begin{marginnote}[]
\entry{Extinction}{The amount of light lost along a single sightline due to absorption or scattering away from the line of sight.}
\entry{Attenuation}{The effective amount of light lost in aggregate for a number of sightlines that includes both the contribution of scattering back into the line of sight, as well as contribution from unobscured stars.}
\end{marginnote}

Extinction represents the amount of light lost along a single line of sight through a dusty medium due to  absorption or scattering away from the line of sight. The observational determination of extinction requires backlights such as stars, gamma ray bursts, quasars, or other objects with much smaller angular extent than a galaxy.
The extinction at a given wavelength results from a combination of the grain size distribution and the optical properties of the grains (which is itself dependent on the chemical composition of the grains). The extinction scales with dust column density.

Although the terms extinction and attenuation are sometimes used interchangeably in the literature, there exist important conceptual differences between the two. In contrast to extinction, attenuation includes the effects arising from the distribution of stars and dust in the galaxy.  Attenuation includes the same mechanisms for loss of photons as extinction, plus scattering back into the line of sight, as well as the contribution to the light by unobscured stars.   Figure~\ref{fig:attenuation_extinction_schematic} schematically summarizes the differences between extinction and attenuation.

\subsection{The Importance of the Attenuation Curve for Recovering Fundamental Physical Properties of Galaxies}
\label{section:importance}

\begin{figure}[h]
\includegraphics[scale=0.4]{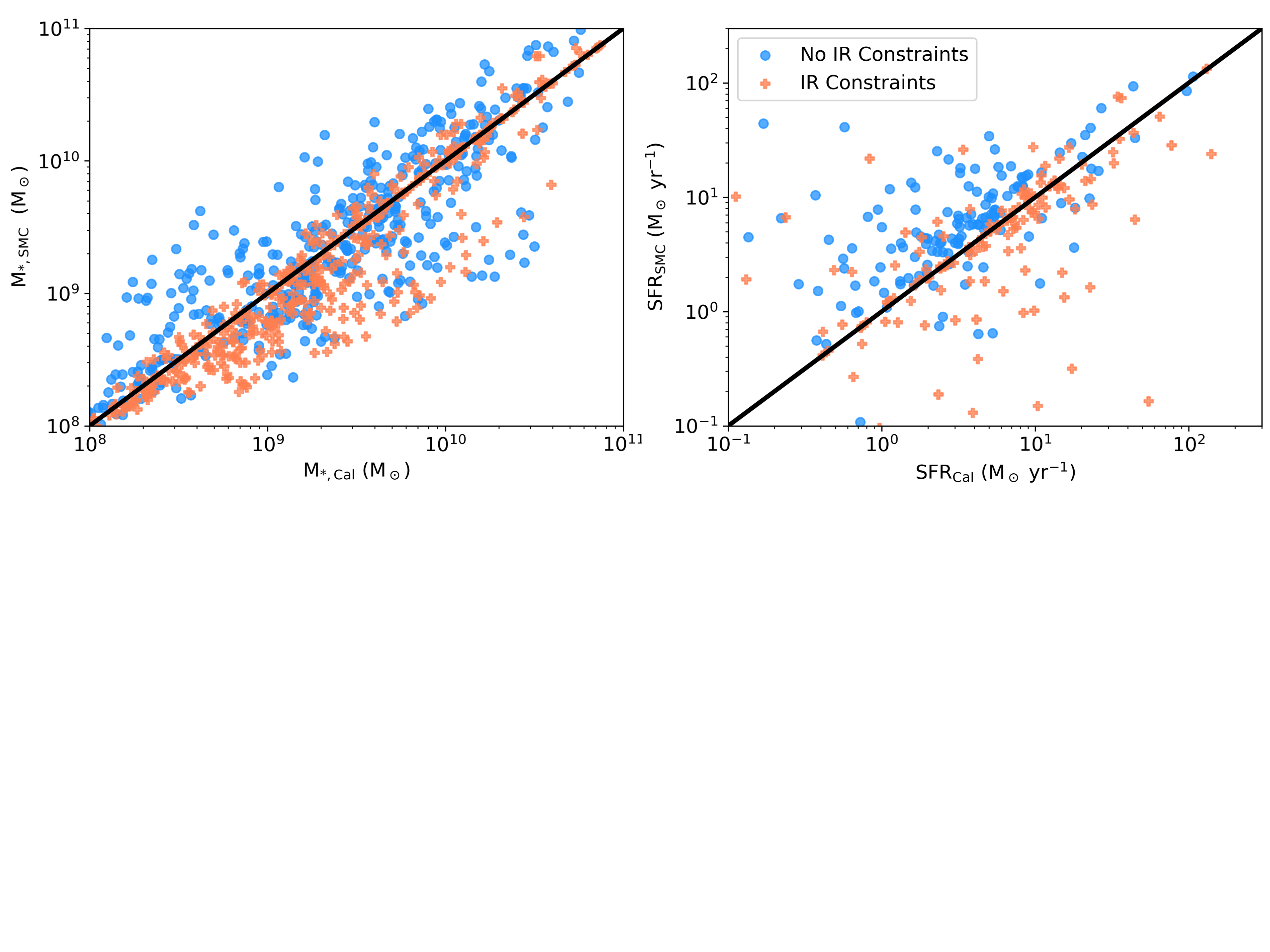}
\vspace{-6cm}
\caption{Results from fitting the mock SEDs of galaxies formed in a cosmological hydrodynamic simulation, but assuming either a \citet{Calzetti2000} attenuation curve in the fits (abscissa), or an SMC extinction curve (ordinate).  Shown are a comparison of derived stellar masses (left) and a comparison of the derived star formation rates (right).  The colors denote whether the SED fit included IR constraints (orange) or not (blue). The right panel excludes fully quenched galaxies. The derived physical properties of galaxies are strongly dependent on the assumed attenuation curve, and highlight the importance of the attenuation curve in modeling the physical properties of observed galaxies.  The calculations for this figure were performed by Sidney Lower (University of Florida), and the authors request that any reproduction of this image attribute credit to her.}
\label{figure:importance}
\end{figure}

To demonstrate the importance of the dust attenuation curve in deriving the physical properties of galaxies, we conduct a simple experiment by generating mock SEDs from $z=0$ galaxies formed in a cosmological hydrodynamic  simulation \citep[the {\sc simba} model;][]{Dave2019}, and then fitting the SEDs with the Bayesian SED fitting software {\sc prospector} \citep{Leja2017,Leja2018,Johnson2019}.  We fit the SEDs using {\it GALEX} and {\it HST} photometry, along with the optional addition of {\it Herschel} IR photometry.  When including the IR photometry, we allow the IR SED to be modeled freely. We assume a $\tau$-model star formation history, and vary only the assumed attenuation curve, employing either a \citet{Calzetti2000} curve, or an average SMC extinction curve. We show the comparison between the derived stellar masses and star formation rates when varying only the assumed attenuation curve in Figure~\ref{figure:importance}.  The derived physical properties can vary dramatically, depending on the assumed dust attenuation curve in SED fitting, a point that has been amplified in the observational literature \citep[e.g.,][]{Kriek2013,Shivaei2015,Reddy2015,Salim2016}.

\input{formalism}

\input{extinction_obs}

\section{\MakeUppercase{The Theory of Dust Extinction Laws}}
\input{extinction_theory}

\section{\MakeUppercase{Attenuation Curves: Theory and Modeling}}
\input{attenuation_theory.tex}

\label{section:attenuation_theory}

\section{\MakeUppercase{Methods of Observationally Deriving Dust Attenuation Curves}}
\input{attenuation_methods.tex}

\section{\MakeUppercase{Parameterizations of Attenuation Curves}}
\input{parameterizations.tex}

\label{section:parameterizations}

\input{atten_obs_lowz.tex}

\input{atten_obs_highz.tex}

\section{\MakeUppercase{Conclusions}}

Large-scale panchromatic and spectroscopic surveys of galaxies, as well as large surveys of Galactic stars, have ushered in a new era in the empirical study of extinction and attenuation laws. Observational advances are further exploited by the development of new analysis and computational techniques, such as stellar population synthesis coupled with Bayesian SED fitting, or the use of standard crayons to study Galactic extinction. On the theoretical side, our understanding of dust grain physics  is improving, while radiative transfer modeling of attenuation in galaxies is becoming more sophisticated and placed in a larger context of galaxy evolution modeling.

\begin{figure}
    \centering
    \includegraphics[scale=0.55]{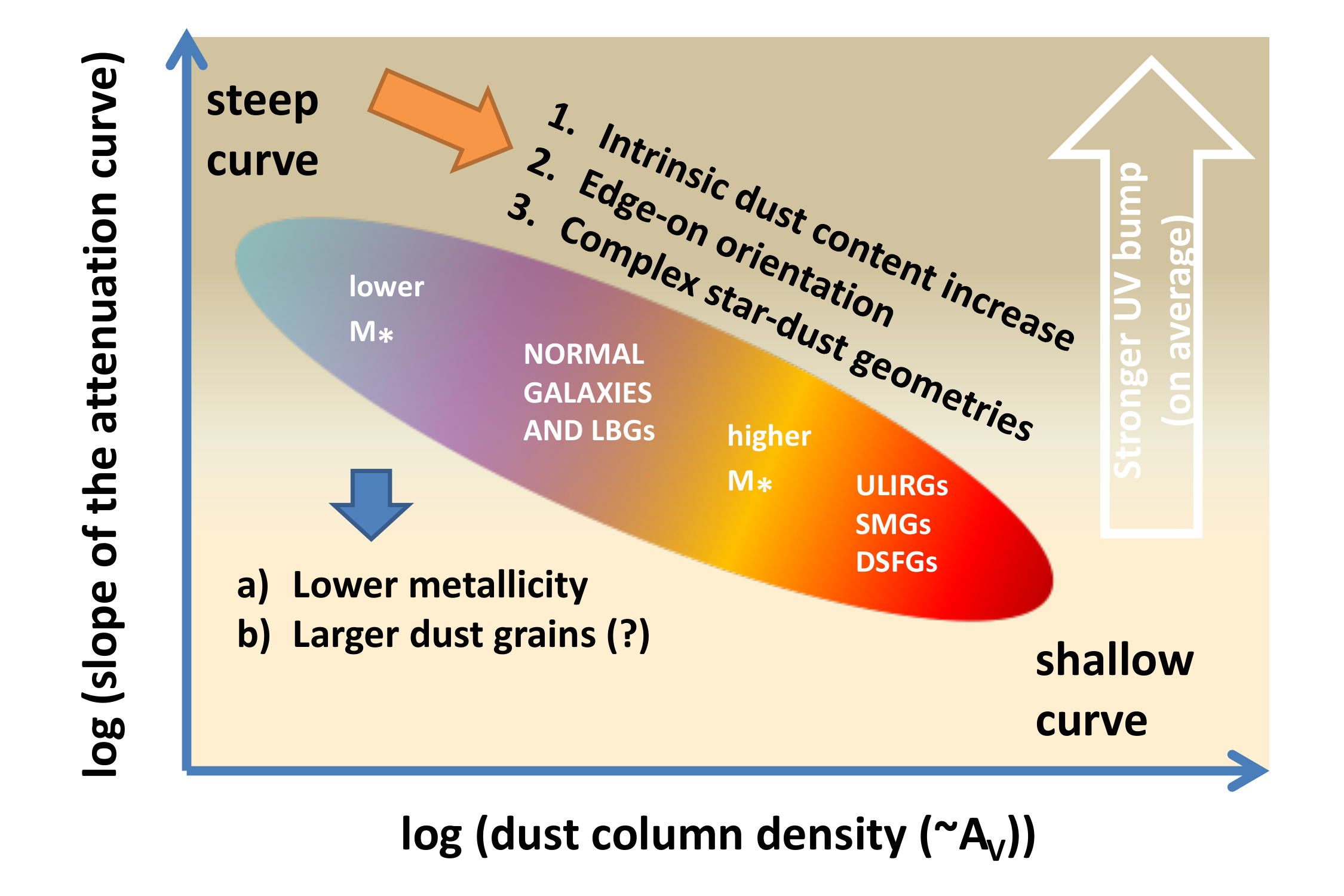}
   \caption{A schematic representation of the dependence of attenuation curve slope on dust column density (for which optical attenuation is a proxy), with possible drivers along the power-law relation (the oval and orange arrow) and possible factors driving the scatter (blue arrow). \label{figure:dal_schematic}}
\end{figure}

We summarize the findings in this review article with  some high-level conclusions regarding extinction and attenuation curves in galaxies.
\begin{itemize}
    \item Local group extinction curves show a great diversity in slopes and UV bump strengths that may form a continuum of properties governed by grain size distribution and composition along with radiative transfer effects dependent on the dust column density and galaxy. This continuum of properties is to be distinguished from a more simplistic discretized split into MW or SMC-type dust.    
    \item A great diversity of attenuation curve slopes (defined as the ratio of attenuation at short and long wavelengths) appears to be present in galaxies at low and high redshifts. For a given amount of optical attenuation, the far-UV attenuation can be $\sim$2--7$\times$ greater, corresponding to a range from shallower than the Calzetti curve to steeper than the SMC curve. Consequently, there is significant scatter between the optical and UV attenuation. This non-universality of attenuation curves is predicted by most radiative transfer models.
    \item Slopes correlate strongly with effective optical opacity, with opaque galaxies having shallower (grayer) curves. The dependence is probably a manifestation of a more fundamental dependence on the dust column density, for which the optical attenuation serves as a proxy. The dependence approximately follows a power law, and is schematically illustrated in Figure \ref{figure:dal_schematic}. Theoretical work informs us that the dependence may be the outcome of a combination of radiative transfer effects (scattering and absorption) and local geometry effects (clumpiness of dust). 
In the local universe, accounting for optical opacity largely removes residual trends in the slope with respect to stellar mass, sSFR or galaxy inclination.     \item The 2175 \AA\ UV bump feature also seems to show a great range in strengths, from effectively non-existent to somewhat above the MW strength (with the bump strength defined as the fraction of peak attenuation at 2175 \AA).     There is a     correlation between the UV bump strength and the attenuation curve slope. Models that reproduce this correlation ascribe bump variations to the infilling from unobscured stars, but intrinsic dust composition differences may also play a part.
    \item The above diversity of curves means that the average curves of galaxies of different stellar masses or SFRs (whose dust contents differ) and of galaxies selected by dust emission (e.g., ULIRGs, DSFGs, SMGs) will span a wide range as well.
\end{itemize}

\begin{issues}[FUTURE ISSUES]
\begin{itemize}
    \item What drives the evolution of the grain size distribution in galaxies across the mass function over cosmic time?
    \item Is there a more unifying framework that explains MW extinction curves alongside those that fall outside of the single-parameter family? 
    \item What is the minimum number of independent parameters necessary to develop a parameterization for attenuation curves that captures the diversity of  effective attenuation curves across different spectral regimes?     \item We should improve the understanding of systematics affecting different observational methodologies in order to resolve the issues where the same samples of galaxies can return different results for attenuation curves. 
      \item What are the potential drivers of attenuation curve slope differences beyond the optical opacity that can be probed observationally? How much of the scatter is due to the different intrinsic extinction curves? How do we test this?
    \item What are the characteristics of attenuation curves in the Lyman continuum region?
           \item Can we use the results from theoretical modeling to better inform SED fitting software (via, e.g., mapping theoretically-derived attenuation curves to observations via learning algorithms).
     
  \end{itemize}
\end{issues}

\section*{DISCLOSURE STATEMENT}
The authors are not aware of any affiliations, memberships, funding, or financial holdings that
might be perceived as affecting the objectivity of this review. 

\section*{ACKNOWLEDGMENTS}
The authors thank Andrew Battisti, V{\'e}ronique Buat, Charlie Conroy, Akio Inoue, Ben Johnson, Mariska Kriek, Joel Leja, Naveen Reddy, Eddie Schlafly, David Schlegel, Kwang-Il Seon, Alice Shapley, Irene Shivaei, S{\'e}bastien Viaene and Thomas Williams for helpful conversations, comments or data.   The authors thank Sidney Lower for her assistance in generating Figure~\ref{figure:importance}, and Bobby Butler for technical assistance and proofreading. S.S.\ wishes to thank Janice C.\ Lee for inspiring discussions over the years, as well as past and current developers of {\sc cigale} SED fitting code, in particular M{\'e}d{\'e}ric Boquien. D.N.\ wishes to express gratitude to Charlie Conroy and Karin Sandstrom for getting him interested in this field to begin with.  D.N.\ was supported in part by NSF AST-1715206 and HST AR-15043.001.

\end{document}

%% file: formalism.tex
\section{Concepts and Formalism of Extinction and Attenuation Curves} \label{section:formalism}

\begin{figure}[h]
\includegraphics[scale=0.55]{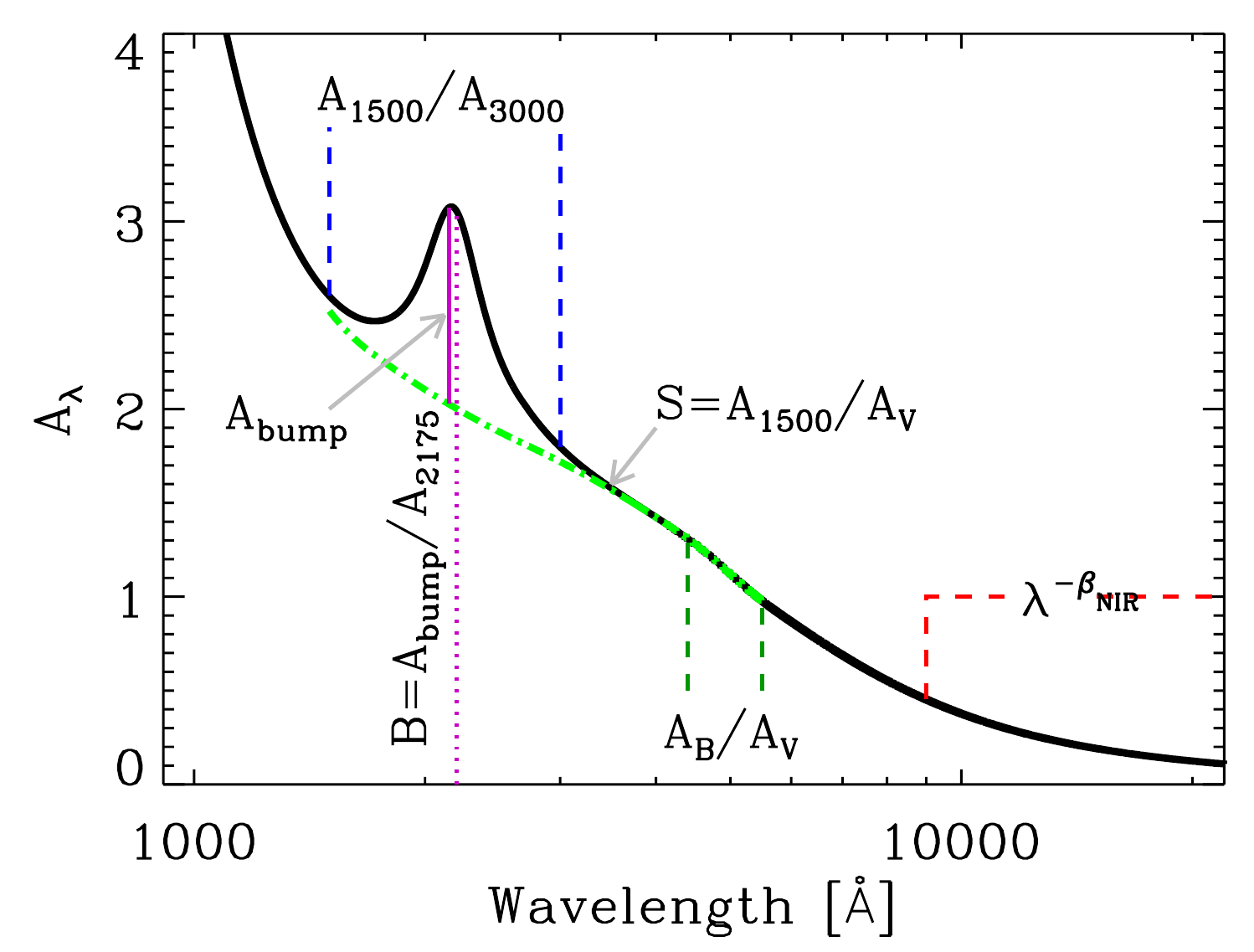}
\caption{Schematic representation of an attenuation or extinction curve
  and the parameters to describe and quantify
  it (we use the \citealt{Fitzpatrick1999} average Galactic extinction curve as our example). The attenuation or extinction curve is the shape of $A_{\lambda}$
    vs.\ wavelength, irrespective of its normalization. Shown here is a
    non-normalized curve, so the values on the $y$ axis are arbitrary. The
    parameters whose definitions are illustrated are the UV
    slope ($A_{\rm 1500}/A_{\rm 3000}$), the strength of the UV bump
    ($B$) , the optical slope ($A_B/A_V$, directly
    proportional to $1/R_V$), and the near-IR power-law
    slope ($\beta_{\rm NIR}$). The UV/optical slope ($S$) is defined over the range shown in green.  
  \label{fig:scheme}}
\end{figure}

There are diverse formalisms and parameterizations that are used to
present and describe dust extinction curves, and, by extension, dust
attenuation curves. In this review we will be using a scheme that
builds upon intuitive concepts, but we will also list some of the
traditional measures or quantities for completeness. In this section,
concepts referring to attenuation and attenuation curves and to
extinction and extinction curves can be used interchangeably.  

If $m_{\lambda,0}$ is an intrinsic, unattenuated spectral energy
distribution (SED) of some astronomical object expressed in
magnitudes, where wavelength $\lambda$ spans a range from the Lyman limit
(912 \AA) to the tail of stellar emission ($\sim 5\mu$m), and
$m_{\lambda}$ is the observed, attenuated SED, then:

\begin{equation}
A_{\lambda} =   m_{\lambda}-m_{\lambda,0}
\end{equation}

\noindent represents the wavelength-dependent attenuation in magnitudes. We can think of $A_{\lambda}$ as consisting of a {\it
  normalization} (e.g., $A_{\lambda}$ at $\lambda\approx 5500$ \AA; 
$A_V$), and a {\it shape} (e.g., $A_{\lambda}/A_V$). It is
the shape of the curve that is
generally referred to as the attenuation/extinction curve/law. Thus, if two galaxies, one with $A_V=0.1$ and another with
$A_V=1$, have the same $A_{\lambda}/A_V$, they are said to have the same
attenuation curve because their shapes are the same. What is different is the normalization, arising
from different column densities of dust.

Curves are often expressed normalized by $A_V$, though the
choice of wavelength or band used for normalization is arbitrary. Older
literature often expressed curves normalized by reddening ($A_B-A_V \equiv E(B-V)$),
and denoted them as $k_{\lambda}$ (meaning that $k_V \equiv R_V$, since $R_V=A_V/(A_V-A_B)$). 
Reddening-normalized curves can be converted into
$A_{\lambda}/A_V$ normalization and vice versa (they contain identical
information), but the interpretation of the former is less intuitive
because the slopes in the optical range (between $B$ and $V$) are
identical for all curves by definition. A more physically motivated but less practical normalization would be by dust column density, or by a proxy, such as the H column density (e.g., \citealt{Mathis1977}).

Detailed individual curves contain a wealth of information that is
difficult to assimilate and compare from one curve to another solely by
looking at the plots of $A_{\lambda}/A_V$ against $\lambda$. It is
therefore useful to define certain parameters of the
curves. In this review we will
focus on two principal and three additional features, whose definitions are shown
schematically in Figure \ref{fig:scheme}, where we use the average Milky Way
extinction curve as an example.  Broadly we define $5$ regimes: the overall UV-optical slope, UV slope, optical slope, near-IR
slope and the UV 2175 \AA\ absorption bump strength. All of these parameters are independent of how
$A_{\lambda}$ is normalized, and are not based on any mathematical parameterization of the curve, which makes them more universal (we will discuss a functional parameterization of attenuation curves in \S~\ref{section:parameterizations}).

\begin{marginnote}[]
\entry{UV-optical Slope}{$S \equiv A_{\rm 1500}/A_{V}$}
\entry{UV (2175 \AA) Bump Strength}{$B \equiv A_{\textrm {bump}}/A_{2175}$}
\entry{UV Slope}{$A_{\rm 1500}/A_{\rm 3000}$}
\entry{Optical Slope}{$A_B/A_V$}
\entry{Near-IR Slope}{$\beta_{\rm NIR}$ where $A_\lambda/A_V = \left(\lambda/5500\AA\right)^{\beta_{\rm NIR}}$}

\end{marginnote}

As a general high-level characterization of a curve, it is useful to define its
overall UV-optical slope, taken to be the ratio of extinctions at 1500 \AA{} and in the $V$-band:
\begin{equation}
  S\equiv A_{1500}/A_V \qquad \textrm {(UV-optical slope)}
\end{equation}
\noindent The shorter wavelength corresponds to the far-UV wavelength
traditionally used as an SFR indicator or in the calculation of the IR excess
(IRX). The attenuation through the {\it GALEX} far-UV bandpass ($A_{\rm FUV}$) is on average 0.03 magnitudes higher than at 1500 \AA. If we assume that attenuation curves are well-modeled by a power-law between 1500 and 5500 \AA, the exponent of the curve $A_{\lambda}/ A_V = (\lambda/5500 \textrm{\AA})^{-n}$ would be
directly related to the UV-optical slope via

\begin{equation}
n = 1.772 \log S
\end{equation}

\noindent though note that the use of $S$ does not imply that the curve follows a power law.  Power-law parameterization requires caution because real curves often depart significantly from a single power-law, so forcing a power-law fit will
be sensitive to the exact range of wavelengths,
sampling, and the treatment of the UV bump. We stress that curves tend to be distributed more uniformly in $n$ (i.e., $\log S$) than in $S$, so an appropriate way to average UV-optical slopes in some sample is using a median or a geometric mean of $S$ values.

The second principal parameter is the strength of the UV bump, which we define as the ratio of extra
extinction due to the bump at 2175 \AA\ ($A_{\rm bump}$; the
  amplitude above the baseline) to the total extinction at 2175 \AA\
  ($A_{2175}$):

\begin{equation}
  B\equiv A_{\rm bump}/ A_{2175}  \qquad \textrm{(UV bump strength)}
\end{equation}

\noindent which is illustrated in Figure \ref{fig:scheme}. The UV bump
strength can formally take a value between 0 (no bump) to 1 (all extinction at
2175 \AA\ is due to the bump). The baseline extinction (i.e., the
extinction in the absence of the bump) can be estimated by
interpolating the extinction shortward and longward of the bump, for example using

\begin{equation}
A_{2175,0} = 0.33 A_{1500}+0.67A_{3000},
\end{equation}

\noindent which we determined empirically from simulated attenuation curves in \citet{Narayanan2018b}. The definition
of the bump used in this review is based on normalizing the bump at the same
grain size (and therefore wavelength), so it is more physically motivated than normalizing
$A_{\rm bump}$ by $A_V$ or by $A_B-A_V$ (the latter is often called $E_b$).  While many literature
parameterizations of the bump also include the width and the central
wavelength of the UV bump \citep[e.g.,][]{Noll2009b,Conroy2010b,Narayanan2018b},
information on these parameters is very limited in the case of
most observed attenuation curves. It has so far proven difficult to relate them to dust
physics even in the case of
extinction curves  \citep{Fitzpatrick1986,Cardelli1989}, so we do not include them among the parameters
discussed here.

In addition to these two principal parameters, for a more detailed study of extinction and attenuation curves it is also useful to define slopes in specific wavelength ranges. We define the UV slope as the ratio of monochromatic extinctions at 1500 \AA \ and 3000 \AA:

\begin{equation}
  A_{1500}/A_{3000}         \qquad \textrm{(UV slope)}
\end{equation}

\noindent where the longer wavelength corresponds to the shortest wavelength not affected by the UV bump. 

We define the optical slope of the curve as:
\begin{equation}
  A_B/A_V \qquad \textrm {(Optical slope)}
\end{equation}

\noindent i.e., the ratio of extinctions in $B$ and $V$
bands, which agrees with the ratio of monochromatic extinctions at 4400 and
5500 \AA\ to within 1.5\% for a wide range of galaxy SEDs. The optical slope is directly related to the
traditional ratio of total to selective extinction ($R_V$), as:

\begin{equation}
A_B/A_V = 1/R_V+1.
\end{equation}

\noindent $A_B/A_V$ is more intuitive than $R_{\rm V}$, with the
added benefit that it is directly related to the normalization of
$A_{\lambda}$ used in the modern literature. 

Following the practice in the literature, the near-IR slope of
the extinction curve is defined as the value of the power-law exponent
$\beta_{\rm NIR}$:

\begin{equation}
A_{\lambda}/ A_V = (\lambda/5500 \textrm{\AA})^{-\beta_{\rm NIR}}  \qquad \textrm{(Near-IR slope)}
\end{equation}

\noindent that best describes the curve in the near-IR wavelength
range, usually starting around 0.9 $\mu$m and extending to 2--5 $\mu$m,
depending on the available data.

%% file: extinction_obs.tex
\section{\MakeUppercase{Extinction curves: Observations and Empirical Properties}} \label{section:ext_obs}

Extinction curves are the building blocks of attenuation curves, and a full understanding of the latter requires understanding of the former. Any variation in the extinction curve between galaxies would also drive variations in the observed attenuation curves. Because the sources of light (stars) and the places where it is
scattered and absorbed (ISM) can have a complex arrangement,
it is nontrivial to map an extinction curve into an effective attenuation curve, and the two differ in ways that are difficult to systematize observationally \citep{Hagen2017,Corre2018}.  Theoretical simulations have shown that, for a fixed extinction curve, the varying dust content and the complexities of the star-dust geometry in galaxies will result in a diverse range of attenuation curves \citep[e.g.,][Section \ref{section:attenuation_theory}]{Narayanan2018b,Trayford2019}.  

\subsection{Methods of Observationally Deriving Extinction Curves}

The best constraints on extinction curves come from the Galaxy and the Magellanic Clouds. Here, we review how the extinction law is determined in these systems; results and
methods regarding more distant galaxies will be covered separately in \S~\ref{section:extinction_extragalactic}. 

Every
method for the determination of extinction curves requires
observations of point sources affected by dust, and their
comparison to reference spectra or SEDs that are dust free. 
The reference spectra can either be empirical or 
 based on theoretical models. There are two principal
approaches, especially when it comes to the study of the Milky Way
extinction curve: the use of individual stars (the ``pair'' method),
and, more recently, the statistical (or survey) method, which uses a
large number of stars.

The pair method has a long history dating back to \citet{Stebbins1939},
who used pairs of B stars to establish that the extinction curve in
the optical region follows roughly a $\lambda^{-1}$ dependence. The pair
method consists of comparing a reddened star with a star of the same
(or nearly the same) spectral type and luminosity class that is
largely dust free.  Stars of type O6-B5 and
luminosity classes III-V are most suitable because of relatively constant intrinsic colors \citep{Fitzpatrick1986}.

If the distances to both stars in a pair were known, and if the
comparison star was truly dust free, obtaining $A_{\lambda}$ in
different bands would be trivial. Since accurate distances in the general case are not known,
the usual approach is to assume that extinction approaches zero at
infinite wavelength. This ``anchoring" will to some extent be
sensitive to the method of extrapolation from the longest
available wavelength to infinite wavelength, so it is
useful to use data extending as far into the NIR as possible while
avoiding potential circumstellar dust
\citep{Savage1979}.

If near-IR data are absent, which was the case in early studies in particular, one can only obtain a {\it relative} extinction curve---extinctions with respect to some band (e.g., $A_{\lambda}-A_V$), derived by comparing the {\it colors} of a reddened and a dust free star, and typically normalized by $A_B-A_V$ (e.g.,
\citealt{Massa1986}). Relative curves are useful for
establishing the characteristics of local features---for example, the
position and the width of the UV bump---but cannot be used
to obtain the parameters required to characterize and compare the curves discussed in Section \ref{section:formalism} .

One limitation of the empirical pair method is that comparison stars often suffer some dust extinction that needs to be corrected \citep{Massa1983}, meaning that the pair method
is of little value to study the extinction curve of low-opacity (high
Galactic latitude) sightlines. This limitation can be overcome by
using theoretical stellar atmosphere models as dust-free
references, which also has the advantage that the extrapolation of near-IR
extinction is not needed \citep{Fitzpatrick2007}.  

Another method to map extinction curves across the sky, including high-latitude sightlines, is to use
a large-survey, statistical approach.  For example, \citet{Peek2010} used red-sequence galaxies from SDSS as ``standard crayons," whose intrinsic colors are known. \citet{Schlafly2010} utilized the fact that the main sequence turn off for field
stars (old populations) has a sharp blue edge, which can be
accurately determined from a large SDSS sample of stars
lying between 1 and 8 kpc. Subsequently, \citet{Schlafly2011} used spectroscopic parameters of SDSS SEGUE stars to predict stellar colors based on models. \citet{Berry2012} used SED fitting to expand the standard crayon method to a general population of stars from SDSS and 2MASS. \citet{Schlafly2014,Schlafly2016} also used optical/near-IR SED fitting, but as a function of spectroscopically determined temperature and metallicity. \citet{Wang2019} used the relatively constant colors of red clump stars to get extinctions.


\subsection{Empirical Properties of Extinction Curves in the Milky Way and Magellanic Clouds}
\label{section:ext_empirical}

\begin{figure}[h]
\includegraphics[scale=0.58]{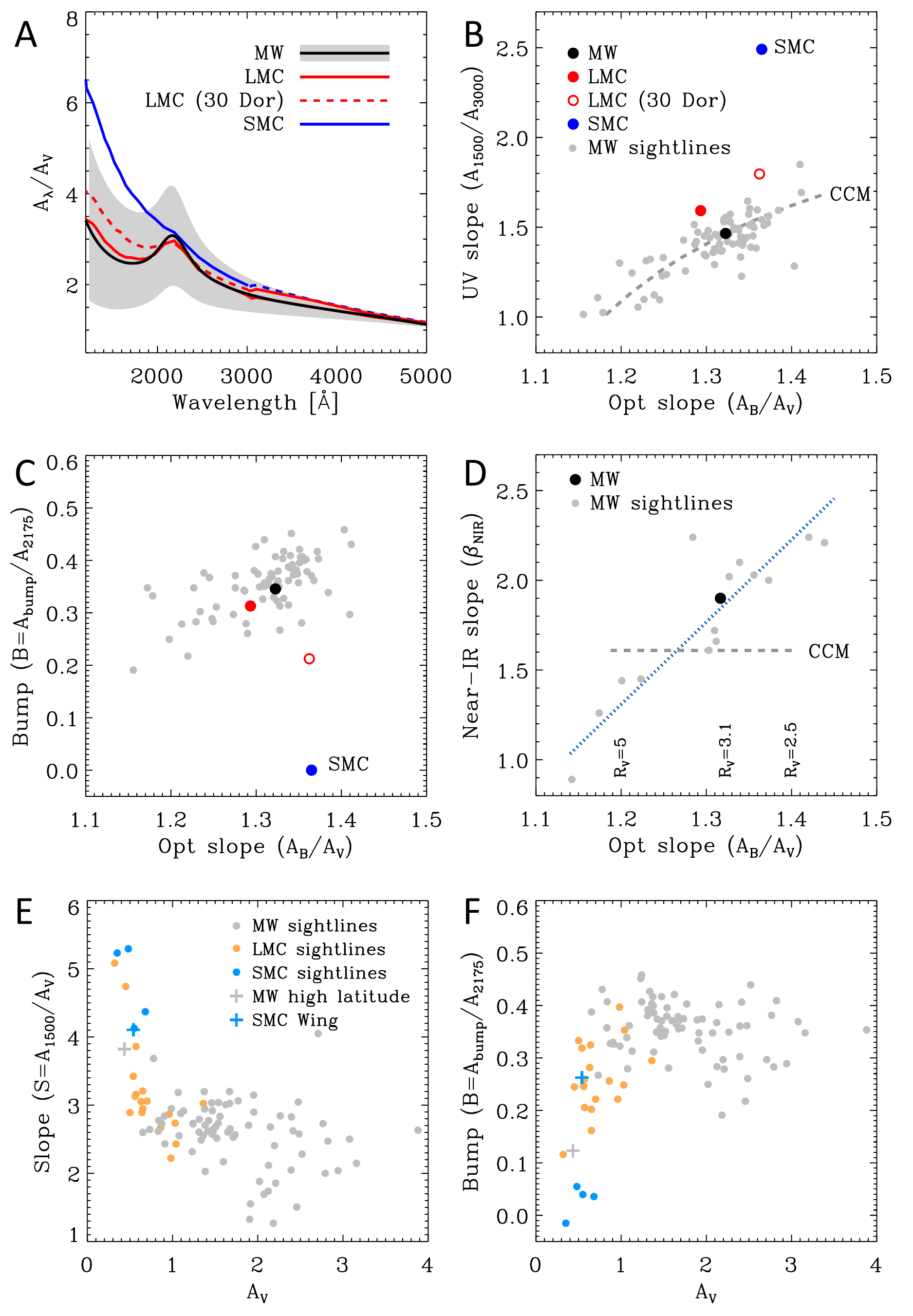}
\caption{The empirical properties of average extinction curves of
  the Milky Way (black), LMC (red) and SMC (blue), as well as of
  individual sightlines within the MW (gray band and dots), LMC (orange), and SMC (light blue).
  Panel A shows $A_V$-normalized extinction curves. Panels B-D show the UV slope, bump strength, and near-IR slope against the optical
  slope, with the CCM relation overplotted in B and D and the linear fit in D. Correspondence between
  $A_B/A_V$ and $R_V$ is shown in D. Panels E and F show the overall slope and bump strength vs.\ optical depth.
  MW data based on \citet{Fitzpatrick1990,Fitzpatrick1999,Clayton2000,Fitzpatrick2007,Fitzpatrick2009,Nataf2016} and SMC/LMC data from \citet{Gordon2003}. 
  \label{fig:ext}}
\end{figure}

In discussing the empirical properties of extinction curves in the Galaxy and Magellanic Clouds, there are two common threads that will emerge: 1)
extinction curves can vary significantly from one line of sight to
another within the same galaxy, and 2) average extinction curves
differ, sometimes dramatically, from one galaxy to another. Point 1
is illustrated in 
Figure \ref{fig:ext}a,
which shows the average Milky Way extinction curve together with a wide band of extinction curves to
individual sightlines in the Milky Way, which all exhibit a UV bump (Section \ref{section:bump}). Point 2 was recognized even in early studies with the {\it International Ultraviolet Explorer (IUE)} that showed that the LMC, and in particular the SMC, have on average steeper and less bumpy curves (Figure \ref{fig:ext}), and has been amplified in the literature over the last decade with major {\it Hubble Space Telescope} campaigns. \citep[e.g.,][]{Sabbi2013,Demarchi2016,Mericajones2017,Romanduval2019}. 


\subsubsection{Extinction Curve Slopes in the UV and Optical}
One of the main questions in the study of extinction is whether
curves to different sightlines in the MW (and more generally)
belong to a {\it single-parameter family} of curves, meaning that
specifying a slope in one spectral region would determine the full curve.

The idea that the slopes over  different wavelength ranges,
from 1200 \AA \ to $\sim$1 $\mu$m, are correlated was first suggested by \citet[][hereafter CCM]{Cardelli1989}. CCM took
relative UV extinction curves towards 31 significantly reddened Galactic plane stars
from \citet{Fitzpatrick1988} and appended them with optical and NIR
photometry to obtain absolute extinction curves. Specifically,
they showed that the slopes between different wavelengths (in
particular the UV wavelengths) and $V$ ($A_{\lambda}/A_V$) are
correlated with the optical slope $R_V^{-1}$ ($\equiv A_B/A_V-1$). Based
on this, they presented an $R_V$-dependent parameterization of curves
that is still widely used, and is referred to by CCM as the ``mean
extinction law'' (for a given optical slope).  The caveat regarding the mean relation, already acknowledged in CCM, is that while different slopes are
correlated with the optical slope, the scatter around the mean
relation at any given value of optical slope is greater than the
observational errors (in particular at $\lambda<1400$ \AA), suggesting
more complex fundamental diversity. 

While the existence of the correlation between UV and optical
extinction curves within the Galaxy is generally accepted, \citet{Fitzpatrick2007}
brought it into question by pointing out that, by anchoring both the UV
and optical slopes at $V$, CCM may have introduced an artificial
correlation even if there was none to begin with. While this concern is valid, it is not clear
that the uncertainties are such to remove the genuine correlation. To
demonstrate this, in Figure
\ref{fig:ext}b we show a pure UV slope defined between two points in the UV
(1500 and 3000 \AA) and not with respect to $V$, against the optical
slope $A_B/A_V$. Data for the $\sim$70 stars come from \citet{Fitzpatrick1990} and
\citet{Fitzpatrick2007}. A relatively narrow
relation emerges, confirming the 
original CCM relation (overplotted in Figure \ref{fig:ext}b). 

However, the single-parameter
family picture that emerged from traditional (low Galactic latitude) MW sightlines does not extend to Magellanic Clouds. Early studies by \citet{Nandy1981, Koornneef1981,Clayton1985} and \citet{Fitzpatrick1985} found that, while there was significant regional variation in the curve across the LMC, on average the LMC had a UV curve steeper than the Milky Way's, especially in the 30 Doradus star-forming complex.  Similarly, early studies of the SMC \citep[e.g.,][]{Prevot1984,Gordon1998} found even steeper UV slopes. In short, while the {\it optical} extinction curve slopes for the LMC and the SMC fall within the range of MW sightlines, the UV slopes tend to be significantly steeper (Figure \ref{fig:ext}b), highlighting the limitation of characterizing curve slopes with  $R_V$ alone. One can ask whether individual lines of sight within the LMC and SMC obey
some other relationship, offset from the MW one, but the current data are not of
sufficient accuracy, nor do the LMC/SMC sightlines have a sufficiently wide range in optical
slopes, to provide a definitive answer \citep{Gordon2003}. From a
practical standpoint, the shape of LMC and SMC curves implies that the CCM $R_V$-dependent mean
curves, or their updates and equivalents \citep{Odonnell1994,Fitzpatrick1999,Valencic2004}), cannot straightforwardly be used for
accurate modeling or parameterization of extinction/attenuation curves in general (see also Section \ref{section:parameterizations}).

\subsubsection{The 2175 \AA\ UV Bump in Local Extinction Curves}
\label{section:bump}
The UV bump \citep{Stetcher1965} represents a
prominent feature in the extinction curves of the MW and LMC, spanning
roughly the wavelength range between 1700 and 2700 \AA. A systematic
study on the observational properties of the bump toward many
sightlines in the MW carried out by \citet{Fitzpatrick1986} revealed that the
peak wavelength of the bump is variable, but the deviations from 2175
\AA \ are relatively small ($1\sigma \approx 5$ \AA). Furthermore, the width of the bump in the Galaxy (FWHM = $480$ \AA, on average) was also
found to be variable, with $1\sigma$
variations of $\approx 20$ \AA. Broad bumps tend to be associated with
denser environments (as assessed visually). The original \citet{Fitzpatrick1986} Galactic study determined that the peak wavelength and the
width of the bump are not correlated with each other, and neither
are correlated with the slope of the extinction curve
\citep{Cardelli1989}.   
 
The feature of the bump that has a greater practical significance is
its strength. \citet{Cardelli1989} showed that the
extra extinction in Galactic sightlines due to the bump, when normalized by $A_V$, is
correlated with the optical slope of the extinction curve ($R_V^{-1}$
or $A_B/A_V$), such that the steeper slopes have stronger bumps. In
Figure \ref{fig:ext}c we revisit this result, but normalizing the bump amplitude to the
extinction at 2175 \AA, which removes any spurious correlation between
the axes. Milky Way sightlines show a spread of a factor of two in UV
bump strength (the average curve has $B=A_{\rm bump}/ A_{2175}=0.35$) and
a correlation with the optical slope is present. Thus the UV bump
fits the scheme of a single-parameter family of extinction curves, at
least in traditional, high-reddening MW sightlines. However, the
LMC 30 Dor average curve has a $\sim$50\% weaker bump than MW
sightlines of the same $A_B/A_V$, whereas the SMC average curve has
essentially no bump \citep{Prevot1984,Gordon1998,Misselt1999,Gordon2003}. Exceptions to this general pattern will be
discussed in Section \ref{section:extinction_parameters}.

\subsubsection{Extinction Curve Slopes in the near IR}
Historically, the properties of the NIR extinction curve have been
more challenging to study than those of the UV and optical curves due to the
lack of data and low extinctions. CCM argued for a 
  universal near-IR extinction law at $\lambda>0.9\ \mu$m, so in their  
parameterization the near-IR power-law slope is fixed to $\beta=1.61$, based on the curve of  \citet{Rieke1985}. This value is no longer favored for the average MW curve. A range of higher values has been found
in various recent studies, as summarized by \citet{Wang2019}, who used a
survey approach to find a typical value of $\beta=2.07$. This value agrees with \citet{Nataf2016}, who found an average $A_V/A_{K_s}=13.4$ ($\beta =1.90$). To reveal any
trends between individual sightlines, \citet{Fitzpatrick2009} used theoretical
stellar atmosphere models to obtain, independent from one another, the
optical and near-IR extinction curves of 14 stars. The power-law
slopes span a range from 0.9 to 2.3, i.e., extending on both sides of
the traditionally used values, and agrees with recent high values at $R_V=3.1$. Interestingly, while the number of data points is small, their near-IR
slopes show a general trend with the optical slopes,  as shown in
Figure \ref{fig:ext}d. Thus, rather than being
universal, the near-IR extinction curve appears to be part of a single
parameter family, at least for the MW sightlines. The shapes of the
near-IR extinction curves of the LMC and SMC are not well known.

\subsection{Extinction Curves and External Parameters}
\label{section:extinction_parameters}
So far we have reviewed and examined correlations between the internal features of
extinction curves, finding trends between traditionally studied MW
sightlines, but not extending to the LMC/SMC average curves. One could ask
if there was an external parameter that could explain the diversity of
curves both within the MW and between the MW and LMC/SMC sightlines, so as
to lead to a continuum of dust properties, as suggested by
\citet{Clayton2000} and \citet{Gordon2003}. In addressing this question we focus on the two features that show the
greatest diversity: the overall UV-optical slope ($S$), and the bump
strength ($B$), and explore how they relate to three potential proxies of dust column
density ($A_{1500}$, $A_V$, $A_B-A_V$), gas column density ($N$(H{\sc i})) and
two variants of the gas-to-dust ratio ($N$(H{\sc i})/$A_V$ and
$N$(H{\sc i})/($A_B-A_V$)). We find the strongest trends against optical
opacity ($A_V$), so in Figures~\ref{fig:ext}e and f, we show individual
sightlines in the MW, LMC (including 30 Dor region) and SMC (LMC/SMC data from \citealt{Gordon2003}).  Particularly noteworthy is the fact that while the LMC and SMC sightlines, on average, exhibit
steeper curves and smaller bumps than the Galaxy, they are also far more transparent than
typical MW sightlines. Remarkably, in regions where the MW and LMC $A_V$ values overlap, the
slopes are similar. Can the gap between the MW and SMC also be
bridged? This question motivated \citet{Clayton2000} to study
high-latitude, low-opacity sightlines in the MW. Their least reddened sightline
($A_V\sim 0.4$, HD 164340), shown as a gray cross in Figure~\ref{fig:ext}e, has a slope steeper than most LMC sightlines and almost
as steep as some SMC ones, plus a nearly absent bump, like in the
SMC. Likewise, the extinction curve toward SMC Wing, a more quiescent and gas-poor sightline, has a
bump as strong as many MW sightlines and a relatively
shallow UV slope (a number of other sightlines in 
the SMC show evidence for a weak UV bump, albeit in attenuation curves; \citealt{Hagen2017}).
Furthermore, the slopes appear to be weakly
correlated with $A_V$ even for high-opacity MW sightlines. Trends similar to those 
described here exist between the UV-only slope
($A_{1500}/A_{3000}$) and $A_V$. On the other hand, the trend is not as pronounced
between the optical slope ($R_V$) and $A_V$, which is consistent with the existence of 
a large range of UV slopes for a given optical slope in non-MW curves, as discussed above.

The general picture that is emerging from both local measurements and those of distant galaxies (Section \ref{section:extinction_extragalactic}) is that extinction curves have a range of steep
slopes and weak bumps for $A_V<0.8$, but are relatively shallow and
bumpy when the $A_V$ (and therefore metallicity) is higher. The dependence on $A_V$ is likely
the consequence of a more fundamental dependence on the dust column
density. 

\subsection{Empirical Properties of Extinction Curves in Other Galaxies}
\label{section:extinction_extragalactic}

The study of dust extinction in extragalactic systems provides an opportunity to understand the physics driving grain size evolution and compositions in galaxies over a diverse range of environments.  
    
\subsubsection{The Local Universe}    

In the local Universe, much of our understanding of extinction curves outside of the Galaxy or Clouds comes from {\it HST}-enabled dust extinction maps in M31 that were derived using  novel color-magnitude diagram fitting methods \citep{Dalcanton2015,Dong2016}.  \citet{Dong2014} measured the extinction curve toward the bulge of M31, and find a curve steeper than the Galactic average ($R_{\rm V} \approx 2.4$--2.5).  Of particular note is that {\it HST} observations of the disk in M31 reveal an extinction curve with comparable slope to the Galactic mean, though with possibly a weaker bump \citep{Bianchi1996}. 
These results, taken together, hint at common trends in the extinction curve in the Milky Way and M31.

\subsubsection{Extinction at High-Redshift}
 Given that the dust extinction law is dependent on grain size distribution and composition, we might indeed expect it to vary not only with environment, but even among galaxies of a similar mass but at different redshifts, as different processes (i.e. growth via metal accretion, coagulation, shattering, or sputtering) may dominate at different times and in different environments.  Similarly, different dust production processes may dominate at early times, and the detections at $z>6$ may place constraints on both formation and growth mechanisms  \citep[e.g.,][]{Draine2009,Riechers2013,Strandet2017,Marrone2018}.  Because of this, there has been significant interest in characterizing the dust extinction law at high redshift. Due to typically sub-solar metallicity in high-redshift galaxies, an SMC-like extinction curve is often invoked.  This said,  as in low-redshift studies, the extinction law is found to have a range of slopes and bump strengths amongst the disparate populations studied over a broad range of cosmic time at high-redshift.

    

Typically, one needs a backlight to produce an absorption spectrum from which to infer an extinction curve.  Common sources include gamma-ray burst afterglows, due to their relatively well-characterized spectra, and quasars \citep[e.g.,][]{York2006,Stratta2007}.  Other groups have taken advantage of magnitude differences at different wavelengths in image pairs of lensed systems to derive extinction laws  \citep[e.g.,][]{Malhotra1997,Falco1999,Munoz2004}, a method developed by \citet{White1992}.

 Of particular interest in high-redshift studies of the extinction curve has been potential variations in the 2175 \AA\ UV bump.  High-$z$ galaxies provide a particularly useful laboratory to study the bump strength owing to the diverse range of metallicities and incident radiation fields.  \citet{Schady2012} studied the mean curve of $17$ GRB host galaxies, and found SMC-like curves with minimal UV bump strengths.  \citet{Zafar2011} used a relatively homogeneous sample of 41 GRB afterglows and found UV bumps to be present in only three cases, which were all associated with $A_V>1$. The general absence of bumps was confirmed more recently in \citet{Zafar2018}, where all GRB extinction curves for which the bump region was observed were featureless (and all had $A_V<0.3$). These same studies have shown that the majority of GRB extinction curves share the steep UV/optical slope of the SMC curve, and like the SMC curve are typically associated with low columns of dust ($A_V<0.5$). Shallow UV/optical slopes similar to the MW were only found in high-$A_V$ sightlines, highlighting the connection between $A_V$ and slopes, suggested by \citet{Schady2012} and discussed in Section \ref{section:extinction_parameters}. Turning to other techniques, \citet{Motta2002} confirmed the presence of a UV bump in a $z\sim 1$ lens galaxy.   In contrast, however, \citet{York2006} find no UV bump and an SMC-like slope in the mean extinction curve of $\sim$800 SDSS quasar absorbers at $1<z<2$. Note that these absorbers tend to have quite low optical depth ($A_V<0.3$). A special search targeting absorbers that exhibit a UV bump showed that such cases do exist. Out of $\sim$30,000 absorbers toward quasars selected in the Sloan Digital Sky Survey at $z \sim 1.0-2.2$, \citet{Ma2015,Ma2017} identified 14 cases in which the UV bump was present, including many that have bumps as large as the Galactic mean. 

\subsection{Correcting for the Milky Way Extinction}

 The Galactic extinction curve is routinely used to correct for the reddening of extragalactic observations. Several open 
questions are related to this procedure. How should we construct maps of reddening ($A_B-A_V$) from which extinctions are derived? Is it appropriate to use an average extinction curve (fixed $R_V$), neglecting the region-to-region
variations? If so, which average curve is the most appropriate, especially in low-$A_V$, high-latitude regions?

Perhaps the most standard dust map in use is the composite {\it COBE/IRAS} map developed by \citet{Schlegel1998}.  These authors constructed a map of dust column densities
from a combination of maps of dust emission at different far-IR
wavelengths, and made corrections using an H{\sc i} map. The dust columns were then converted into reddening ($A_B-A_V$) using a calibration sample
of $\sim$ 400 elliptical galaxies, whose intrinsic $B-V$ colors were inferred
from the dust-insensitive Mg$_2$ spectroscopic index
\citep{Faber1989}. It should be
pointed out that in the case of a curve varying across the sky, the \citet{Schlegel1998} map may more closely follow extinction at 1
$\mu$m than $A_B-A_V$ \citep{Schlafly2010}. This may explain the temperature-related local departures in the \citet{Schlegel1998} reddening map with respect to $A_B-A_V$ determined independently using a statistical method \citep{Peek2010}. Furthermore, \citet{Schlafly2011} used another statistical method to derive reddening maps in SDSS areas, finding an overall normalization different by 14\% from \citet{Schlegel1998}. Reddening maps were expanded in coverage to 3/4 of the sky by using colors of Pan-STARRS1 stars \citep{Schlafly2014}, which also allowed finer resolution compared to the \citet{Schlegel1998} map. Another high-resolution reddening map was produced using H{\sc i} data by \citet{Lenz2017}, but covering 40\% of the sky. 

\citet{Schlafly2010} found the optical slope of the extinction curve to
vary across the high-latitude sky ($2.7<R_V<3.4$, i.e., $1.29<A_B/A_V<1.37$), but much less 
than the traditional Galactic plane sightlines using the pair method, thus justifying the use of a universal extinction curve. \citet{Berry2012} found a very similar range in slopes when combining optical and near-IR constraints. Interestingly, a similarly small range of variation in $R_V$ ($\sigma=0.2$) was found along the Galactic plane \citep{Schlafly2016}, where the reddening is typically ten times higher than at high latitudes. Furthermore, \citet{Schlafly2016} find that optical slopes are correlated with the dust emissivity, though the origin of the correlation remains unknown.


\citet{Schlafly2010} and \citet{Schlafly2011} found that the \citet{Fitzpatrick1999} extinction
curve with $R_V=3.1$ slope agrees better with their optical constraints than either CCM or its update by \citet{Odonnell1994} (with any $R_V$). However, at longer wavelengths, \citet{Schlafly2016} find that the \citet{Fitzpatrick2009} near-IR curve is a better choice than any other literature parameterization, presumably because of its $R_V$-dependent shape (Figure \ref{fig:ext}d). When it comes to the UV, obtaining an accurate high-latitude extinction curve is more difficult due to the presence of the bump and shallower data from the {\it GALEX} satellite. \citet{Yuan2013} found that the far-UV extinction is 2/3 of near-UV extinction, in contrast to standard extinction laws which predict similar extinction in the two bandpasses. In contrast, using colors of galaxies, \citet{Peek2013} found that far-UV and near-UV extinction are as much as 50\% and 25\% higher than the \citet{Fitzpatrick1999} extinction, which would make the high-latitude Galactic extinction curve more similar to the LMC 30 Dor curve and therefore not part of the MW single-parameter family. This finding supports the idea that the UV/optical slopes of extinction curves may care more about the column density than in which galaxy they are measured.

Reddening maps continue to improve as well as our understanding of the variation of the extinction curve across the sky. Extinction curves agree well with the standard curves in optical range, but more work is needed to understand the departures in the UV at high Galactic latitudes.

%% file: extinction_theory.tex
In this section, we synthesize broader trends in the theory of interstellar extinction and modeling efforts.  
For a comprehensive review of the relevant equations, we point the
reader to the excellent Saas-Fee notes by \citet{Draine2004}.

\subsection{Extinction Law Theory: Basic Framework}
\label{section:extinction_theory}

The optical depth seen by photons interacting with dust is $\tau_{\rm
  dust}\left(\lambda\right) = N_{\rm d}Q_{\rm ext}\left(\lambda\right)\pi a^2$, where $N_{\rm d}$ is the column
density of dust, $a$ is the size of the grain (therefore assuming a
spherical geometry) and $Q_{\rm ext}$ is the ratio of the extinction
cross section to the geometric cross section.  The extinction cross
section is the sum of the absorption and scattering cross sections, and depends on the grain composition \citep{Draine2011}.  In order to
develop a theoretical model for the extinction law, one must 
therefore simultaneously develop a model for the size
distribution ($a$), as well as for the
composition of the interstellar grains.

Most theories are constrained by a variety of observed emission and
absorption features associated with dust.  The existence of the 2175
\AA\ extinction bump \citep[e.g.,][]{Stetcher1965} points to at
least some carbonaceous grains in the ISM \citep[owing to
  corresponding absorption features in $sp^2$-bonded carbon
  sheets;][]{Draine2011}.  This connection is amplified by extinction
features at $\sim$3.4 $\mu$m that are likely due to carbon-based
grains \citep[e.g.,][]{Duley1983}.  At the same time, absorption
and emission near 9.7 $\mu$m and 18 $\mu$m indicate the likely
existence of silicates as an important constituent of interstellar dust
\citep[e.g.,][]{Gillett1975,Kemper2004}.  
These constraints have motivated two major types of
modeling efforts: synthesis models and direct numerical simulation.
In what follows, we summarize progress in each area.

\subsection{Extinction Law Theory: Synthesis Models}

The essence of synthesis modeling is the development of a theory that
simultaneously models the grain size distribution and dust composition,
while remaining constrained by (1) the observed
wavelength-dependent extinction in the ISM, (2) polarization as a
function of wavelength, (3) atomic depletion patterns in the ISM, and
(4) the infrared emission properties of galaxies.  One of the first
synthesis models, and indeed the quintessential study in this
category, was performed by \citet[][hereafter MRN]{Mathis1977}, who utilized a non-parametric size distribution, and solved for the best-fit grain size distribution.  MRN found that a power-law size distribution of spherical grains  of graphite-silicate composition \citep[see also][]{Hoyle1969} provided a reasonable model for the observed extinction law at the time. 

The MRN model was extended by \citet{Draine1984}, both in wavelength
coverage and in the optical properties of the graphites and silicates.  
Further updates into the X-ray regime (i.e., extended dielectric functions for graphites, silicates, and silicon carbides) were developed
by \citet{Laor1993}.  Subsequent models have departed from the
 power-law grain size distribution. For example, \citet{Kim1994}
utilized a ``maximum entropy'' method for deriving the size
distribution. 
Similarly, \citet{Li2001} characterized the size distribution of very small grains by multiple log-normal distribution functions.

The last few decades have seen significant updates to the
graphite-silicate model for dust.  The first major class
of models were an extension to include a polycyclic
aromatic hydrocarbon (PAH) component
\citep[e.g.,][]{Siebenmorgen1992,
Dwek1997, Li2001, Weingartner2001, Siebenmorgen2014}, which typically
describe PAHs as carbonaceous grains, with the mass fraction divided
into PAHs vs.\ graphites as a free parameter.  Generally, in these
models the small PAH molecules contain $<10^3$ C atoms
and have physical and chemical properties similar to those of larger
graphite-like grains, though in reality there are some significant 
differences arising from the C-H bonds in PAHs. \citet{Weingartner2001} derived a size
distribution of grains to match observed extinction patterns in the Galaxy, LMC, and SMC, and included PAHs such that $\sim$15\% of the
interstellar C abundance was locked up in PAHs.  A significant step
forward was taken by \citet{Draine2007}, who leveraged the
results from the {\it Spitzer Space Telescope} to derive a size distribution
of PAHs that reproduces the average Milky Way emission
spectrum \citep[though note that the absorption cross section of PAHs
remains a significant uncertainty; e.g.,][]{Siebenmorgen2014}.  Other
models in the graphite-silicate family include amorphous carbons or
composite particles consisting of a mixture of silicates,
refractory organic materials, and water ice
\citep[e.g.,][]{Zubko2004,Galliano2011}, and models with silicates with
organic refractory mantles combined with very small carbonaceous
particles and possibly PAHs \citep[e.g.,][]{Desert1990,Li1997}.  

These models largely treat interstellar dust as a mixture of
different constituents, meaning that while the 2175 \AA{} carrier drives FUV extinction, often a separate species is also included to contribute to the FUV extinction when the 2175 \AA{} bump is not present.  An alternative class of models,
monikered the ``core-mantle'' models, sought to unify these populations by treating dust grains as a silicate core with a carbonaceous mantle
that varies in composition or depth depending on the
environment \citep[e.g.,][]{Greenberg1986,Duley1987,Jones1990,Li1997,Ysard2016}.
Recently, \citet{Jones2017} have developed the {\sc themis} framework,
which provides a model for a core-mantle grain that varies in
composition and/or mantle depth based on the response to the local
environment (e.g., volumetric/surface densities, radiation field).  

\subsection{Extinction Law Theory: Numerical Models}

In recent years, computational advances have permitted the direct
simulation of the evolution of dust grains (in aggregate) in galaxy
evolution simulations, which serves as a complement to synthesis models in
understanding extinction in galaxies. The relevant physical processes governing
the total dust mass are formation in stellar ejecta, growth by
accretion of gas-phase metals, and destruction by sputtering in hot
gas, while the grain-size distribution is primarily governed by
shattering and coagulation processes.  In practice the processes are
actually linked, as growth rates may depend on the surface area of the
grain \citep{Kuo2012}.  Of course, the problem must be initialized: a
grain size distribution from different production processes (i.e., SNe
or AGB stars) is typically assumed as a boundary condition.

A significant step forward in the numerical modeling of the evolution
of grain sizes was made by \citet{Asano2013} and \citet{Nozawa2015}, who
aggregated the relevent equations for the aforementioned processes and embedded them in a simplified (one-zone) galaxy evolution
model, demonstrating the feasibility of a self-consistent galactic
chemical evolution model wherein the size distribution of grains could
be directly simulated.  These models were extended by
\citet{Hirashita2015}, who developed a ``two-size'' approximation as a
means for reducing the computational load of dynamically modeling a
broad size spectrum of dust grains in bona fide hydrodynamic galaxy
evolution simulations.  Here, grains are divided into large 
(typically $a>0.03$ $\mu$m) and small grains ($a<0.03$ $\mu$m).  This
model has been adopted by a number of groups since then
\citep[e.g.,][]{Gjergo2018,Aoyama2018,Hou2019}, and has been shown in
at least some cases to reproduce the results from multi-grain models
when the large and small grains represent
log-normal distributions \citep{Hirashita2015}.  

Multiple grain sizes have been implemented in idealized hydrodynamic
galaxy evolution simulations \citep{McKinnon2018,Aoyama2018}. By combining these size distributions with assumed extinction
efficiencies from the literature, \citet{Aoyama2018},
\citet{McKinnon2018}, and \citet{Hou2019} extended their numerical
simulations of extinction curves in galaxies.  Both groups find the physics of coagulation and shattering to be critical in
reproducing both the FUV rise and 2175 \AA\ bump in Milky
Way-like extinction curves.

The current status of galaxy formation simulations is to include
either single-size grains
\citep[e.g.,][]{McKinnon2016,Zhukovska2016,Popping2017,Dave2019,Li2019,Vijayan2019} or
two grain sizes \citep[e.g.,][]{Hirashita2015} into cosmological galaxy
formation simulations, or multiple grain sizes into idealized
simulations \citep[e.g.,][]{Asano2013,McKinnon2018}. In the next few
years, we anticipate numerical models progressing significantly through the implementation of a full size distribution of grains in cosmological
hydrodynamic simulations on the fly to develop  cosmological models for
the grain size evolution in galaxies as they evolve across cosmic time.





%% file: attenuation_theory.tex
We now turn our attention from extinction to attenuation, and begin by providing an overview of the status of theoretical models aimed at understanding variations in the shapes of attenuation laws.   We first outline the landscape of different methods, which vary in spatial scale from the regions surrounding individual star clusters to cosmological volumes.  We then aggregate the key results that thread through different models, attempting to synthesize major themes that have emerged over the last few decades in this area.

\subsection{Basic Concepts}

Fundamentally, all models deriving attenuation laws share the same
basic methodology: (1) they develop a model for the structure of
galaxies and populate them with stars and ISM; (2) they make assumptions regarding the extinction
properties of the dust grains in those galaxies; and (3) they model the radiative
transfer of stellar light as it escapes these systems.  The major
difference between models lies in how the star-dust geometry
is specified. Simulation methods range from relatively simplified analytic
prescriptions for galaxy structure to complex hydrodynamic galaxy
formation simulations.  More complex simulations
likely reflect a more realistic star-dust geometry, though at the
expense of both increased computation time and ease in
isolating the impact of individual physical phenomena on the final
dust attenuation curve.

Analytic models were pioneered by \citet{Witt1996} and \citet{Witt2000}, and have been employed
by a number of groups in subsequent years
\citep[e.g.][]{Gordon1997,Charlot2000,Pierini2004,Inoue2005,Fischera2011}.  These models were expanded
upon by \citet{Seon2016}, who developed a model for a $3$D turbulent ISM
patch within a galaxy, described by a clumpy geometry and a log-normal gas density
distribution function.  Other groups have developed analytic
models for the $1$-dimensional radial structure of dust in galaxies in
efforts to model the dust attenuation curve
\citep[e.g.,][]{Inoue2006,Popescu2011}.  The structure of the ISM in analytic models can range from simple screen models \citep[e.g.,][]{Calzetti2000} to models that include stellar birthclouds \citep[e.g.,][]{Charlot2000,Wild2011}.

Increasing in complexity, a number of groups have turned to
semi-analytic models (SAMs) in order to derive the attenuation
properties of a diverse population of galaxies
\citep[e.g.,][]{Granato2000,Fontanot2009,Fontanot2011,Wilkins2012,GonzalezPerez2013,Popping2017b}.
In the context of galaxy formation simulations, SAMs are a family of
models that draw on a combination of either analytic or numerical
models for halo growth over cosmic time, with analytic prescriptions
for the evolution of the baryonic content and structure within halos \citep[see the overview by][]{Somerville2015}.  The majority of SAMs employ a relatively simplified galaxy
structure (for example, a stellar bulge embedded in an axisymmetric
disk of gas, dust, and stars).  While the geometry of galaxies in SAMs
is still analytically pre-specified, SAMs build on the complexity of the
aforementioned static models in that the stellar population content
(i.e., age and radial distribution) are cosmologically motivated.

Hydrodynamic galaxy evolution models offer increased sophistication in
their computed star-dust geometries, though this costs compute
cycles.  The highest resolution hydrodynamic models are idealized
galaxy simulations which take place not in a cosmological context, but
rather by evolving galaxies that are initialized with a simplified
geometry \citep[e.g.,][]{Jonsson2006,Rocha2008,Natale2015}.  In
hydrodynamic models, the star-dust geometry can vary over time due to secular
dynamical processes, as well as the redistribution of gas and dust due
to stellar/AGN winds, though the effects of cosmic accretion and
structural changes that owe to hierarchical merging will be missed.
These issues are mitigated by large-box cosmological simulations (or
cosmological ``zoom in'' simulations), which model individual galaxies in their native cosmic environment
\citep[e.g.,][]{Narayanan2018b,Trayford2019}.

\subsection{Theoretical Drivers of Attenuation Law Slope Variation}
A general source of consensus amongst theoretical models of attenuation laws is the role of the star-dust geometry in galaxies, as well as the impact of dust column density.   The former are broadly subdivided into two regimes:  geometry on the scale of birthclouds surrounding star clusters, and the large-scale geometry of diffuse dust in the ISM.  We do not discuss the global geometry (i.e., viewing angle), as this is simply a manifestation of the aforementioned effects.

\subsubsection{The Role of Birthclouds}
\label{section:birthclouds}
\citet{Fanelli1988} found that the attenuation toward nebular lines in their sample of galaxies was significantly higher than that of the stellar continuum. The result was confirmed by \citet{Calzetti1994}, who found the typical ratio of $A_B-A_V$ reddening to be $\sim$2. These results motivated the introduction of a two-component model of dust attenuation by \citet{Charlot2000}. In this model, the radiation from all stars is subject to attenuation by a diffuse dust (ISM) component, but stars below a threshold age $t_{\rm threshold}$, corresponding physically to the dispersal time of the birthcloud ($\sim$10 Myr; \citealt{Blitz1980}), see an additional source of attenuation by way of a natal birthcloud.

The \citet{Charlot2000} model assumes that birthclouds and the ISM each attenuate light according to fixed power-law attenuation curves, which we refer to as the {\it component} curves, to distinguish them from the resulting effective attenuation curves:
\renewcommand{\arraystretch}{2}
\begin{equation}
A_{\lambda} = \left\{ \begin{array}{cl} A_{V,{\rm ISM}}\left(\frac{\lambda}{\lambda_{V}}\right)^{-n_{\rm ISM}}+ A_{V,{\rm BC}}\left(\frac{\lambda}{\lambda_{V}}\right)^{-n_{\rm BC}} & t\leq t_{\rm threshold}\\
A_{V,{\rm ISM}}\left(\frac{\lambda}{\lambda_{V}}\right)^{-n_{\rm ISM}}&t>t_{\rm threshold}\end{array}\right.\,\,,
\end{equation}
where $n_{\rm BC}$ and $n_{\rm ISM}$ correspond to power-law exponents for the diffuse ISM and natal birthclouds.  \citet{Charlot2000} found that a single power-law exponent ($n_{\rm ISM} = n_{\rm BC} = 0.7$), though with different normalizations for the optical depths of birthclouds ($A_{V,{\rm BC}}$) and diffuse dust ($A_{V,{\rm ISM}}$), was able to sufficiently satisfy both the nebular line constraints as well as the \citet{Meurer1999} relationship between the infrared excess and ultraviolet slope (IRX-$\beta$) in nearby galaxies (see  Section \ref{section:lowz_irxb} for more details). The slope of the birthcloud curve was later argued to be $n_{\rm BC} = 1.3$ in \citet{daCunha2008}, and this value is used in the SED fitting package {\sc magphys}. 

The effective attenuation curve resulting from the \citet{Charlot2000} model evolves with time (becoming steeper) following a strong burst of star formation. Furthermore, \citet{Inoue2005} showed that because
young luminous stars suffer from extra attenuation and also dominate at short wavelengths, the resulting effective curve will be steeper than the component curves. In the \citet{Inoue2005} analytical model, the effect of the clumpy geometry on the attenuation curve arise primarily because high-density regions contain young luminous stars.

The two-component model has shown significant utility in the nearly two decades since its publication.
For example, \citet{Trayford2019} showed that hydrodynamic galaxy formation models require the inclusion of a birthcloud component to match the observed optical depth-attenuation curve slope relation in galaxies (discussed further in \S~\ref{section:theory_slopes}).  Similarly, the birthcloud model has allowed for additional degrees of freedom in SED fitting software, including both age-selective attenuation, as well as the ability to model birthclouds with a range of parameterized curves, and not just the power-law curves as in the original model \citep[e.g., {\sc cigale}][]{Noll2009}.

\subsubsection{The Relationship between Geometry and the Slope of the Attenuation Law}
\label{section:theory_slopes}

Theoretical models are largely in consensus about the relationship between the geometry of the star-dust mixture in galaxies and the slope of the 
dust attenuation law.  In short, more simplified
geometries with obscuration in the ultraviolet result in steeper
curves, whereas more complex geometries with clumpy dust result in
flatter curves.

This result was first demonstrated in the seminal series by
\citet{Witt1996,Gordon1997} and \citet{Witt2000}, who used simplifed
analytic geometries to highlight the impact of the star-dust geometry
on the slope of the attenuation curve.  These authors found that, as a
general rule, as the inhomogeneity increased in the ISM structure,
attenuation curves become greyer (flatter).  This result has been
amplified in the literature by both analytic models
\citep[e.g.,][]{Ferrara1999,Bianchi2000,Charlot2000,Inoue2005,Fischera2011,Wild2011}, and hydrodynamic
or similar methods
\citep{Natale2015,Seon2016,Trayford2017,Narayanan2018b,Trayford2019}.  The general physical driver behind greyer attenuation curves with more
complex star-dust geometries is that UV photons from massive stars
more easily escape, and reduce the net measured optical depths at
short wavelengths.  \citet{Narayanan2018b} showed that a similar
effect can transpire for normalized attenuation curves with increasing
fractions of unobscured old stars as well, underscoring the concept
that more complex geometries generally result in greyer curves. 

\subsubsection{The Relationship between Dust Content and the Slope of the Attenuation Law}
Another important driver of the attenuation curve slope is the total dust column density along the line of sight. As observed by (e.g.,) \citet[][]{Salmon2016} and \citet{Salim2018}, larger optical depths $A_V$ correspond to grayer (shallower) attenuation curves.  Many theoretical studies have predicted this effect \citep{Witt2000,Pierini2004,Inoue2005}, but it was in particular highlighted by \citet{Chevallard2013}, who aggregated and analyzed a diverse series of theoretical attenuation law studies by \citet{Pierini2004,Tuffs2004,Silva1998} and \citet{Jonsson2006}, and showed that all the studies predict, with some normalization differences, a relationship between the optical depth $A_V$ and attenuation law slope.  This relationship was subsequently postdicted in hydrodynamic cosmological zoom simulations \citep{Narayanan2018b}, as well as in large-box cosmological hydrodynamic simulations \citep{Trayford2019}. A physical origin for the optical depth-slope relationship was posited by \citet{Chevallard2013}.   While the scattering albedo used by most theoretical studies is roughly constant in the optical, the phase functions are not: red light scatters more isotropically than blue.  Red photons emitted in the equatorial plane of the galaxy will be more likely to escape, while blue photons have a comparatively increased likelihood of remaining, and subsequently being absorbed.  In the low optical depth limit, this corresponds to a steepening of the attenuation curve.  In the high $A_V$ limit, the contribution to the UV is dominated by stars located outside the $\tau=1$ surface, and the shape of the curve is flattened by both scattering into the line of sight as well as radiation from unobscured OB stars \citep{Narayanan2018b}.  

\subsection{The Origin of 2175 \AA\ Bump Variations in Attenuation Curves}
\label{section:theory_bump}
When simulating bump strength variations in attenuation curves,
theoretical models differ fundamentally on one central point: is it
possible to create a bump-free attenuation curve when the underlying extinction
curve has a bump in it? Posed another way, does the 2175 \AA \ bump strength in the attenuation curve tell us anything about the intrinsic bump strength in the extinction curve?

The first class of models requires either removing the bump from the underlying extinction curve, or significantly reducing it in order to produce a bump-free attenuation curve.  For example, early models by \citet{Gordon1997} and \citet{Witt2000} that employed
simplified analytic geometries found that a silicate-dominant grain
composition (i.e. ``SMC-like'' dust) that does not have a bump was
necessary to produce bump-free attenuation curves.  Similarly,
\citet{Fischera2011} found, using turbulent slab models, that
destroying the 2175 \AA\ carriers in the underlying extinction curve is the most efficient means of producing a bump-free curve. 
More recent high-resolution models of a
turbulent ISM by \citet{Seon2016} have found that radiative transfer effects can work to reduce the bump strength in galaxies, though a reduced bump amplitude (by $\sim$60--70\%) in the underlying 
\citet{Weingartner2001} dust model is necessary to produce fully bump-free curves.  
At the same time, a number of recent works have suggested that even
galaxies whose dust properties allow for a bump can exhibit bump-free
attenuation curves, either owing to the particulars of the star-dust
geometry or albedo effects \citep[e.g.,][]{Seon2016}.  This is in contrast to studies that require a bump-free extinction curve to produce a bump-free attenuation curve.  For example,
\citet{Narayanan2018b} used a combination of dust radiative transfer
calculations with cosmological zoom-in simulations of galaxy evolution
to demonstrate that geometry effects alone can remove the bump from a
galaxy's attenuation curve, even when using a traditional \citet{Weingartner2001}
size distribution and a \citet{Draine2007} composition (i.e. one that
includes a bump).  In this situation, UV emission from unobscured
massive stars fills in the bump absorption feature.  Similarly, \citet{Granato2000} and
\citet{Panuzzo2007} utilized semi-analytic models to show that a
complex star-dust geometry can also fill in the UV bump, though in
contrast to \citet{Narayanan2018b} found that it was
unobscured old stars providing the extra UV emission as the massive
stars by construction were obscured by birthclouds.

A strong constraint on models for bump variations was introduced by \citet{Kriek2013}, who found that steeper attenuation curves for high-redshift galaxies have stronger bumps (Section \ref{section:atten_highz}).  \citet{Seon2016} and \citet{Narayanan2018b} found that radiative transfer effects and clumpier media tend to flatten attenuation curves, as well as reduce bump strengths.  In this class of models, then, a relationship between the bump strength and tilt of the curve is natural.  Unfortunately, both models that employ reduced bump strengths in the underlying extinction curve and those that do not are able to match the \citet{Kriek2013} observed bump-tilt relation, thus limiting its utility in distinguishing between models.

\begin{figure}
    \centering
    \includegraphics[scale=0.75]{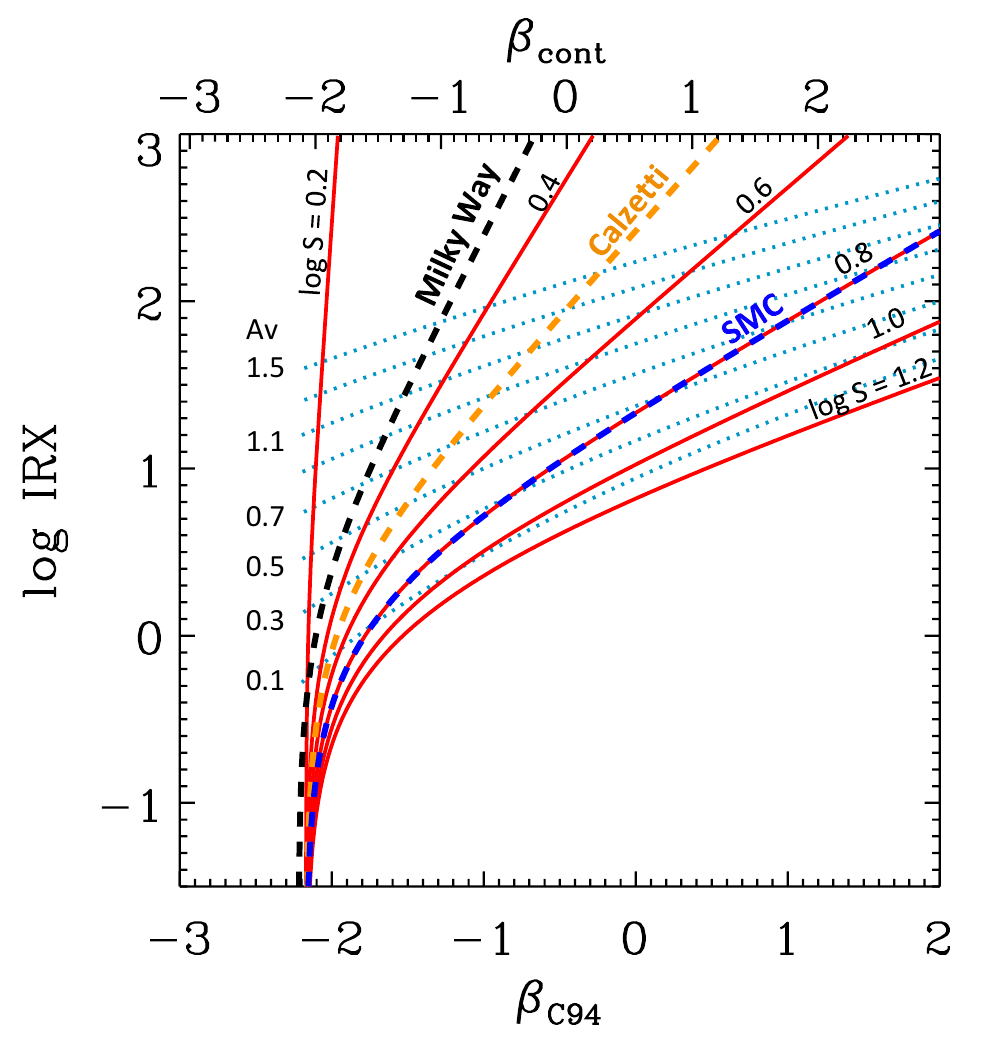}
   \caption {The IRX-$\beta$ diagram showing the loci of galaxies with different attenuation curve slopes $S$ (including MW and SMC extinction and the Calzetti attenuation curves) and  effective optical attenuations $A_V$. Curves are obtained from stellar population synthesis models attenuated using a \citet{Noll2009} parameterization and restricted to the parameter space occupied by actual star-forming galaxies. The scatter in the IRX-$\beta$ plane at fixed attenuation is primarily the result of the varying attenuation curves of galaxies. The UV slope is defined using the \citet{Calzetti1994} windows, and also converted to a continuous window value on the alternative axis. \label{figure:irxb_attenuation}}
\end{figure}

\subsection{Theory of the IRX-$\beta$ Relation}
\label{section:irxb_theory}

The dust attenuation properties in galaxies are often characterized in terms of the relationship between their infrared excess (IRX) and UV continuum slopes ($\beta$). The IRX is defined as:
\begin{equation}
    {\rm IRX}\equiv \frac{L_{\rm IR}}{L_{\rm FUV}}
\end{equation}
\noindent where $L_{\rm IR}$ is the total IR luminosity, while the FUV luminosity is usually defined through {\it GALEX} FUV filter, or as monochromatic luminosity at 1500 \AA\ or 1600 \AA. The UV continuum slope, $\beta$, is defined as the index in the power-law relationship $f_\lambda \propto \lambda^\beta$ over the wavelengths that lie in the 1200--3000 \AA\ range (we discuss the observational methodology in Section \ref{section:irx_method}).  A relationship between the two for a fixed attenuation curve is natural:  as ultraviolet energy is transferred into the infrared through dust reprocessing, the UV power-law index $\beta$ should redden.  Furthermore, because for star-forming galaxies with realistic star-formation histories and metallicities the intrinsic, dust-free UV slope varies very little \citep{Leitherer1999,Calzetti2001}, IRX-$\beta$ can be used to probe the shape of the attenuation curve. This concept is demonstrated in Figure~\ref{figure:irxb_attenuation}, where the loci of galaxies with different attenuation curve slopes (attenuated using the \citealt{Noll2009} parameterization; Section \ref{section:parameterizations}) are shown alongside the MW and SMC extinction curves and the \citet{Calzetti2000} attenuation curve.

A wide range of theoretical methodologies have been employed to understand expectations for the observed IRX-$\beta$ relation in galaxies.  These include analytic models \citep[][]{Ferrara2017,Popping2017}, semi-analytic models \citep[e.g.,][]{Granato2000}, idealized galaxy evolution simulations \citep{Safarzadeh2017}, and cosmological hydrodynamic galaxy formation simulations \citep{Mancini2016,Cullen2017,Narayanan2018a}. For the most part, there is consensus amongst the varied theoretical efforts. \citet{Granato2000} demonstrated the importance of decoupled sites of emission of UV and infrared radiation via consideration of the impact of birthcloud clearing times on their simulations.  Similarly, \citet{Narayanan2018a} explicitly calculated the UV optical depths for hydrodynamic simulations of galaxy evolution to arrive at a similar point, while \citet{Popping2017} cast this result in the context of holes in obscuring dust screens, as well as the impact of increased turbulence in the obscuring ISM.     In short, increasing optical depths move galaxies along the IRX-$\beta$ relation (from bottom-left to top-right in Figure~\ref{figure:irxb_attenuation}), while increasingly decoupled star-dust geometries (i.e. a proxy for varying attenuation curves) displace galaxies vertically in IRX-$\beta$ space.  
That the scatter in where star-forming galaxies present in IRX-$\beta$ space is due to the diversity of underlying attenuation laws is a point of consensus amongst theoretical models, and has indeed also been suggested in the interpretation of various observational datasets \citep[e.g.,][]{Boquien2009,Boquien2012,Mao2014,Salim2019,Alvarez2019}.   What is unknown, however, is the root cause of the  attenuation law variations.  There is a degeneracy between variations in dust extinction curves (i.e., variations in grain sizes and/or the optical properties of grains) and the star-dust geometry in driving attenuation law variations.  Both can impact the location of galaxies in the IRX-$\beta$ space by modifying the attenuation curve.

\begin{figure}
    \centering
    \includegraphics[scale=0.67]{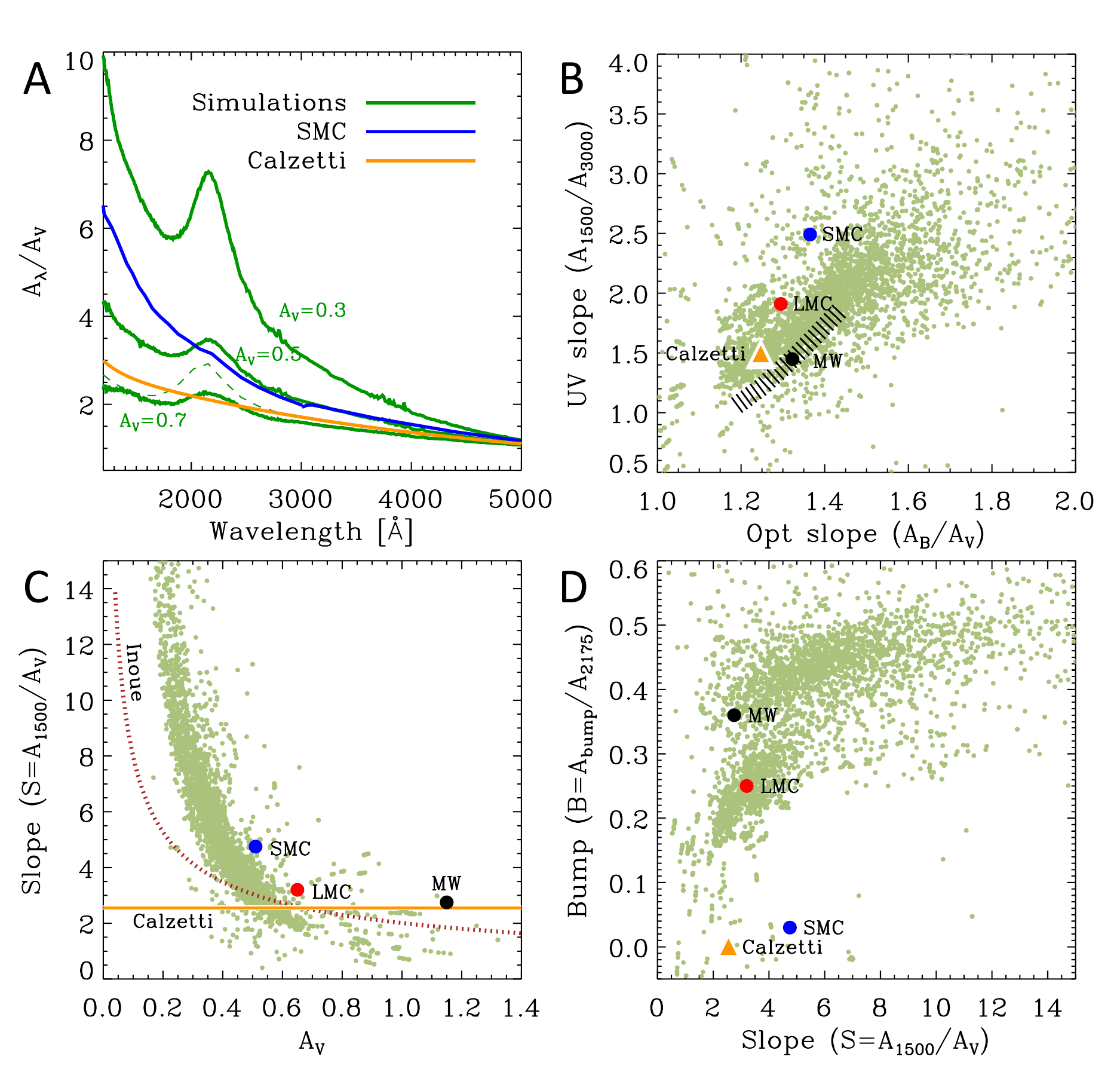}
    \caption{Predicted attenuation curves from hydrodynamic cosmological simulations coupled with radiative transfer modeling. Panel A shows median simulated curves in several bins of optical opacity, alongside the Calzetti and SMC curves as comparisons. The dashed line represents the input extinction curve (of the Milky Way) used in simulations. Panel B shows the UV vs.\ optical slopes, indicating that there is a correlation, though much wider than the single-parameter family of MW extinction curves (black stripes). Panel C shows a very strong correlation of the UV/optical slope on optical attenuation, also predicted using analytical models of \citet{Inoue2005} (dotted line), and panel D shows how the UV bump strength correlates with the UV/optical slope. \label{figure:sim}}
\end{figure}

\subsection{Modeling Predictions}

We now briefly show, by example, the relationship between extinction and attenuation in theoretical simulations by creating a plot (Figure \ref{figure:sim}) analogous to Figure~\ref{fig:ext}, i.e., relating the slopes of attenuation curves to one another. It will be evident that even though extinction curves along a number of sightlines in the Galaxy occupy a relatively narrow swath in these slope-slope and slope-opacity spaces, attenuation curves that use Milky Way-like extinction curves underpinning the grain physics will produce significantly more scatter, owing entirely to the complexities of the star-dust geometry and varying dust column densities. 

For this exercise we utilize the cosmological hydrodynamic {\sc mufasa} simulations presented in \citet{Narayanan2018b}, which were run in a ($25$ Mpc)$^3$ box with $512^3$ particles.  These were coupled with stellar population synthesis models and $3$D dust radiative transfer simulations \citep{Narayanan2015} that determine the attenuation of the stellar radiation.  The models presented here include galaxies at $z=0$ and $z=2$. The input extinction curve in the simulation is fixed at the MW curve given by \citet{Weingartner2001}, and shown in Figure \ref{figure:sim}a as a dashed line. The \citet{Calzetti1994} attenuation and SMC extinction curves are shown for reference, along with median curves (medians of $A_{\lambda}/A_V$ at each $\lambda$) from the simulations, binned according to different optical opacities. Figure \ref{figure:sim}b shows simulation predictions for the relationship between the UV and optical slopes. The plot extends to much steeper UV and optical slopes than the equivalent plot for the extinction curves (Figure \ref{fig:ext}b). There is a correlation between the UV and optical slopes in a sense similar to the trend seen in individual extinction sightlines in the Galaxy (black stripes), though it is roughly $2\times$ wider than the extinction relation.  This is a consequence of the star-dust geometry giving a wide range of attenuation laws, even for fixed grain size distribution and composition.

Moving to the overall UV/optical slope ($S=A_{1500}/A_V$), which is easier to constrain observationally than the UV or optical slope alone, we show its relationship with the optical opacity in Figure \ref{figure:sim}c.  The simulations help to elucidate a trend such that more opaque galaxies have shallow curves. Specifically, galaxies with $A_V>0.6$ are expected to have slopes shallower than the Calzetti curve. Remarkably, galaxies with $A_V<0.4$ are expected to have steeper curves than even the SMC. The slopes exponentially, with the highest values corresponding to the power-law index of 2.0. A scatter around the trend is $\sim$0.12 in $\log S$. We also show the results (assuming SMC extinction curve) of the analytical two-component model of \citet{Inoue2005}, which employs a fixed clumpy geometry, wherein young stars are only found in dense clumps. A higher fraction of young stars in dense regions may be responsible for shift towards lower $A_V$.  Finally, as pointed out in Section \ref{section:ext_obs}, the mean extinction curves for MW, LMC and SMC appear to follow a qualitatively similar trend between the slope of the curve and the mean opacity of sightlines.

In Figure~\ref{figure:sim}d we turn our focus to the relationship between the predicted bump strength and the slope of the attenuation curve.  The bump strength is correlated with the slope,  which may reveal the impact of star-dust geometry on bump strength, in that the reduced bump strength appear to correspond to decreased UV opacity, consistent with the explanation provided earlier in this section.

%% file: attenuation_methods.tex
Broadly, in order to derive the attenuation curve for a galaxy, one must be able to infer the unattenuated SED.  The methods for deriving the dust attenuation curve can be divided into two categories based on whether this unattenuated (or less attenuated) galaxy SED is inferred empirically or is established based on the stellar population synthesis (SPS) models.  Methods can also be characterized based on whether they derive attenuation curves for ensembles of galaxies in aggregate, or aim to do so for individual galaxies.

\subsection{Empirical Comparison Methods}
\label{section:templates}

For the empirical methods it is important to classify or rank galaxies by attenuation in order to use less attenuated galaxies as references. Ideally, the comparison galaxies should be without dust and have the same intrinsic SED as dusty galaxies, but in practice the  dust content of comparison galaxies can be non-zero so long as their (relative) attenuation can be accurately determined. 

\citet{Calzetti1994} pioneered work on attenuation curves with the introduction of a {\it template} empirical method, which they applied to derive the attenuation curve of local starbursting galaxies. The method first produced a relative attenuation curve ($A_{\lambda}-A_V$), which was then anchored with the IR data \citep{Calzetti2000} to get the absolute curve ($A_{\lambda}/A_V$). The data, consisting of UV/optical spectra,  were binned into $6$ groups, based on the value of their Balmer decrements.
The Balmer decrement measures the attenuation of H{\sc ii} regions surrounding massive stars and is used in this method as a proxy for the stellar continuum attenuation, on the basis of its correlation with the steepness of the UV slope ($\beta$), which for continuously star-forming galaxies will mostly vary as a result of dust attenuation, and less so due to the variations in mean population age or stellar metallicities \citep{Calzetti1994}. The spectra in each group of galaxies are averaged (after normalizing by the flux at 5500 \AA) to produce template spectra. The template based on the galaxies with the smallest Balmer decrement is then used as a reference spectrum, to which other templates are compared to obtain relative attenuation curves ($\delta A_{\lambda}-\delta A_V$). These curves are normalized by the relative attenuation between the templates from the Balmer decrement, converted to stellar attenuation using a fixed factor, and then averaged to obtain the final relative attenuation curve. The method of Calzetti has been adapted by \citet{Battisti2016} to use UV photometry instead of spectra. 

An absolute attenuation at some wavelength is required to go from a relative attenuation curve to a true attenuation curve. As in the case of an extinction curve, this anchoring can be accomplished by assuming that attenuation approaches zero at longer wavelengths \citep{Calzetti1997b,Battisti2017a}, but the result will depend on the exact method of extrapolating to infinite wavelength \citep{Reddy2015}. In the case of attenuation curves, anchoring can also be achieved using dust emission, by requiring that luminosity absorbed by the dust in stellar continuum matches the luminosity emitted by the dust, i.e., the total IR luminosity \citep{Calzetti2000}. In principle, the template method can be adapted to use the IRX instead of the Balmer decrement. In that case the absolute attenuation curve ($A_{\lambda}/A_V$) can, in principle, be determined without separate anchoring. An IRX-based empirical method is used by \citet{Johnson2007}, who compare colors of galaxies of similar age but different IRX to derive aggregate attenuation curves. 

A variant of the template method is the pair method of \citet{Kinney1994}, in which individual galaxies are compared instead of the templates. In this method, the galaxies are assumed to be intrinsically similar, but with different effective dust attenuations. \citet{Wild2011} adapted this method to large samples in the SDSS, and constructed attenuation curves by averaging a large number of pair comparisons, in each of which one member is significantly more attenuated than the other as judged by their Balmer decrements. The galaxies in the pair are chosen to have similar redshifts, specific SFRs, metallicities, and inclinations. Observed SEDs are anchored in the $K$-band, where the attenuation is relatively small, so no separate anchoring is needed as in the case of \citet{Calzetti1994}. Importantly, in the pair method the Balmer decrements are not used explicitly to derive attenuation, but only to determine which galaxy is more attenuated.

In certain fortuitous cases where two galaxies overlap, the attenuation curve of a foreground galaxy (its overlapping part) can be obtained using an empirical method originally developed by \citet{White1992}, in which the background galaxy serves as the backlight with SED known from the non-overlapping portion \citep[e.g.][]{Berlind1997,Elmegreen2001,Holwerda2009,Keel2014}. A similar method can be used to derive the attenuation curves of resolved dust lanes in early-type galaxies \citep{Viaene2017}.

The principal advantage of empirical methods is that they do not require a parameterization, allowing the shapes of the curves or their features to be studied in greater detail, especially with spectroscopic data (e.g., \citealt{Noll2007,Noll2009b}). Template-based empirical methods do not derive the slopes of attenuation curves of individual galaxies. Their main disadvantage is the implicit assumption that individual curves are relatively homogeneous and on average similar between different templates or between the members of a pair. If that is not the case, and in particular, if the slope of the curve depends on the dust content itself, the curves characteristic of the dustier template/pair member will carry a greater weight in the final average curve \citep{Wild2011}. More generally, if individual curves span a large range of slopes, then there is no single attenuation measure that allows ranking (e.g., $A_{1500,1}>A_{1500,2}$ does not imply $A_{V,1}>A_{V,2}$).

\subsection{Model-based Comparison Methods}

 Theory predicts that attenuation curves should vary dramatically from galaxy to galaxy \citep[e.g.,][Section \ref{section:attenuation_theory}]{Narayanan2018b}. To quote the \citet{Conroy2013} review on SED fitting techniques, ``the use of a single dust attenuation curve for analyzing a wide range of SED types is [...] without theoretical justification.'' Nevertheless, for a long time SED fitting techniques were performed with fixed attenuation curves. A fixed-curve approach is implemented partially because of computational ease (allowing a well-sampled free curve increases the computation time significantly), and also because of the perceived futility of constraining the attenuation curve due to the well-known dust-age-metallicity degeneracy. Recently, however, it has been shown that useful constraints on the curve are possible even without IR data \citep{Kriek2013,Salim2016,Salmon2016}.  As a result, the SED fitting technique is gaining momentum as a method for the determination of attenuation curves, regardless of the availability of dust emission constraints.

Currently two popular SED fitting packages, {\sc cigale} \citep{Boquien2019} and {\sc prospector} \citep{Johnson2019,Leja2017}, allow attenuation curves to be derived in a parametric form. SED fitting techniques involve the employment of SPS models in the creation of a library or grid of dust-free galaxy SEDs, spanning a range of star formation histories and stellar metallicities (given a fixed stellar initial mass function, isochrones, and atmosphere models). These SEDs are then attenuated according to some dust prescription to yield observed SEDs, from which the model observed-frame photometry is extracted. The fitting is performed next, in which observed photometry is compared to model photometry with known galaxy parameters (the mass-to-light ratio, specific SFR, ages, attenuation, etc.). SED fitting techniques for galaxies were traditionally used to identify a single best-fitting model, whose galaxy parameters were then adopted as the nominal values (e.g., \citealt{Spinrad1971,Faber1972,Sawicki1998,Papovich2001}). The assessment of errors in the best-fitting approach could be achieved by re-fitting with perturbed input photometry (e.g., \citealt{Wiklind2008}). Subsequently, Bayesian fitting techniques have been developed  \citep{Kauffmann2003,Salim2005,Salim2007,Noll2009,Leja2017,Johnson2019}, in which posterior probability distribution functions (PDFs) for every galaxy parameter of interest (or their combination) are derived. Adopting the mean or the median of a PDF as the nominal value results in more robust estimates than  a single most likely model, and the width of a PDF provides an estimate of parameter errors.

Significant improvements in the estimation of $A_{\lambda}$ using the SED fitting method emerge if constraints from dust emission are included \citep{Burgarella2005}. SED fitting techniques that include infrared photometry are typically referred to as energy-balance methods, because the IR SED is normalized in models in such a way that the total IR luminosity from stellar-heated dust emission matches the luminosity absorbed by the dust in UV through near-IR. In energy-balance SED fitting, the shape of the IR SED is described by parameters which, depending on the IR model, range from a single parameter \citep{Dale2002} to six parameters \citep{daCunha2008}. The addition of parameters increases model library sizes and fitting computation time if the entire SED is fit concurrently. However, unless stellar continuum properties are used to constrain the shape of the IR SED (e.g., through the specific star formation rate; \citealt{daCunha2008}), the IR SED fitting is essentially decoupled from the UV/optical/near-IR fitting. This has led to an alternative approach, SED+LIR fitting, in which the IR luminosity is determined separately, and then used as a direct energy-balance constraint in the UV-near-IR SED fitting \citep{Salim2018}. Furthermore, in cases when there is only a single reliable flux point, or the IR SED is not well-sampled in the far-IR, the latter approach allows the IR luminosity to be reasonably well determined from luminosity-dependent IR templates (e.g., \citealt{Chary2001}) and then used in the SED+LIR fitting. Template-based IR luminosities can be refined using empirical corrections constructed from galaxies that have full IR sampling \citep{Salim2018}.

Using SED fitting to constrain the dust attenuation curve or other galaxy properties is a powerful technique if employed appropriately. First, the model colors must cover the parameter space of observed colors. If that is not the case, the ranges of model quantities (the ``priors'') need to be expanded or adjusted. Ensuring a good match typically requires the inclusion of emission lines in the models or their exclusion from broad-band photometry \citep{Kauffmann2003,Salim2016}. If a match cannot be achieved, it is best to exclude the bands that cannot be properly modeled since they may bias the fits of good bands. 
Similarly, the systematic photometry errors need to be chosen adequately to yield the expected best-fit $\chi^2$ distribution of the ensemble, as well as the expected distribution of individual photometry band residuals. 

The dust attenuation curve in the SED fitting can be implemented directly as an effective curve, or using the \citet{Charlot2000} two-component model (see Section \ref{section:birthclouds}).  Note that the attenuation curve pertaining to birthclouds is very poorly constrained in SED fitting \citep[e.g.][]{Conroy2010b,LoFaro2017}, so it is justified to keep it fixed in SED fitting, or to tie it to the slope of the ISM dust attenuation curve. In the latter case the resulting effective curve will be different (steeper) than the input (component) curves because younger populations that dominate at shorter wavelengths suffer higher attenuation \citep{Inoue2005}.

An additional uncertainly can arise when modeling objects that may include an active galactic nucleus (AGN).  Specifically, when including IR data, care must be taken to exclude the contribution of AGN heating to the infrared luminosity, which can be significant among objects detected in surveys that detect only the most luminous IR objects at some redshift \citep{Daddi2007}.  If the IR SED is well sampled, the AGN contribution can potentially be mitigated by including it in the modeling (e.g., \citealt{Leja2018}).

Finally, we note that although the modeling approach for deriving attenuation curves has mostly been employed using SED fitting techniques, there are some exceptions. \citet{Conroy2010b} use models to predict UV and optical colors and compare them with aggregate colors of galaxy with different inclinations. \citet{Scoville2015} identify a sample of intrinsically similar galaxies with ongoing SF (as evidenced by C{\sc iv} absorption) and young mean ages (a small Balmer break) for which they derive an average attenuation curve by comparing the SEDs to model (dust-free) starburst spectra. \citet{Conroy2010c} observes the change in colors of galaxies as a function of redshift and compares it to model predictions to constrain the UV bump strength as it traverses the observed wavelengths. Finally, some studies derive an attenuation curve as a part of detailed modeling of a galaxy SED using radiative transfer calculations (e.g., \citealt{DeLooze2014}).

\subsection{The IRX-$\beta$ Method}
\label{section:irx_method}

The IRX-$\beta$ method can be considered an offshoot of model-based methods, having specifics that justify separate treatment. The IRX-$\beta$ diagram relates the IRX to the power-law exponent $\beta$ of the observed UV SED slope \citep{Meurer1995}. For star-forming galaxies, IRX is tightly correlated with $A_{\rm FUV}$ or $A_{1500}$. \citet{Meurer1999} showed that local starbursting galaxies (overlapping with the sample of \citealt{Calzetti1994}) occupy a relatively narrow locus in the diagram, forming an IRX-$\beta$ relation. The potential existence of a relation was significant because it laid the foundation for a straightforward correction of UV emission in galaxies where IR observations are lacking, based on the more easily observable UV slope (or color).  

The observed UV slopes of star-forming galaxies range from $\beta=-2.5$ to $\beta=2$. However, the SPS models predict the {\it intrinsic} (dust-free) UV slopes to span a range that is 10 times smaller, $-2.5<\beta_0<-2.1$, even when galaxies with extremely high sSFRs in the early universe are considered \citep{Cullen2017}. Consequently, a dust-corrected IRX-$\beta$ diagram would consist of a narrow pillar of points. To more clearly illustrate the point, let us assume that the horizontal width of the pillar in IRX-$\beta$ space is zero. Reddening these points with a {\it fixed} attenuation curve will shift them to the right, in a way that preserves the tight relation. This is because a given IRX in combination with a fixed curve can yield only a unique value for the observed $\beta$ (or UV color). Therefore, in the case of a small spread of intrinsic UV colors, the scatter in the observed IRX-$\beta$ beta plane must result from a diversity of attenuation curves in the UV. Steep IRX-$\beta$ slopes would correspond to shallow attenuation curves and vice versa.  We show this explicitly in Figure \ref{figure:irxb_attenuation}. 

For a given attenuation curve parameterization (Section \ref{section:parameterizations}) and assumed bump strength, one can therefore construct a mapping between the values of IRX and $\beta$ on one hand, and the values of attenuation curve slope and $A_V$ on another. Following \citet{Salim2019}, one can parameterize the IRX-$\beta$ plane in a way that directly translates to the slope of the attenuation curve:
\begin{equation}
    \log S = a_0\frac{\beta-\beta_{\rm min}}{\log({\rm IRX}/{\rm BC}+1)}+a_1
\label{eqn:irx}
\end{equation}

\noindent We use minimization of scatter of low-redshift galaxies from \citet{Salim2019} to find $\beta_{\rm min}=-2.17$, BC$=1.65$, in excellent agreement with the \citet{Overzier2011} values for local starbursts. The parameters assume that the UV slope $\beta$ is determined over 10 spectral windows, following \citet{Calzetti1994}. Other definitions can yield significantly different values of $\beta$. For example, the slope from fitting over the entire SED continuum from 1268 to 2580 \AA\ ($\beta_{\rm cont}$) is related to $\beta$ from Calzetti et al.\ as $\beta_{\rm cont} = 0.55+1.21 \beta_{\rm C94}$ \citep{Salim2019}. The mapping between the slope of the attenuation curve and IRX-$\beta$ is subject to some degree of degeneracy because the UV bump can affect $\beta$ \citep{Mao2014}, modifying the coefficients in Equation \ref{eqn:irx} and introducing noise. The degeneracy can be alleviated using the fact that the bump strength and the UV slope in real galaxies are to some degree correlated \citep{Kriek2013}. Therefore, using the above sample of low-redshift galaxies and assuming that attenuation curves follow the modified Calzetti curve parameterization (Section \ref{section:parameterizations}), we obtain $a_0 = 0.42$ and $a_1=0.08$. The slope $S$ is recovered with an RMS of 0.08 dex. The intrinsic UV slope shifts towards the blue as a function of the specific SFR relatively slowly ($d\beta_0/(d\log {\rm sSFR})=-0.13$), suggesting that Equation \ref{eqn:irx} with locally determined coefficients and a small shift in $\beta_{\rm min}$ would be applicable at higher redshifts. For a given attenuation curve parameterization, $A_V$ can also be derived from the IRX-$\beta$ diagram, as shown by the tracks in Figure \ref{figure:irxb_attenuation}.

The scatter in the IRX-$\beta$ plane can thus be used to probe the diversity of attenuation curve slopes. This connection has been explicitly exploited in a number of studies (e.g., \citealt{Salmon2016,Cullen2017,LoFaro2017,McLure2018,Reddy2018,Alvarez2019}), and implicitly in many more. However, in practice, the IRX-$\beta$ method has several issues. First, it is difficult to measure the UV slope without being affected by the 2175 \AA\ UV bump (e.g., \citealt{Buat2011b,Tress2018}). Even if the bump is removed or accounted for, the estimate of $\beta$ will depend on the choice and width of filters, the redshift, and the wavelength range over which the slope is fit. Furthermore, though the effect is smaller in comparison with that of the UV bump degeneracy, the intrinsic UV slope (which is dependent on the choice of isochrones and atmosphere models) will depend on the sSFR (or mean age), potentially complicating comparisons at different redshifts \citep{Reddy2018}. Finally, the IRX-$\beta$ plane primarily probes the UV component of attenuation curves \citep{LoFaro2017}.

%% file: parameterizations.tex
\begin{figure}
    \centering
    \includegraphics[scale=0.67]{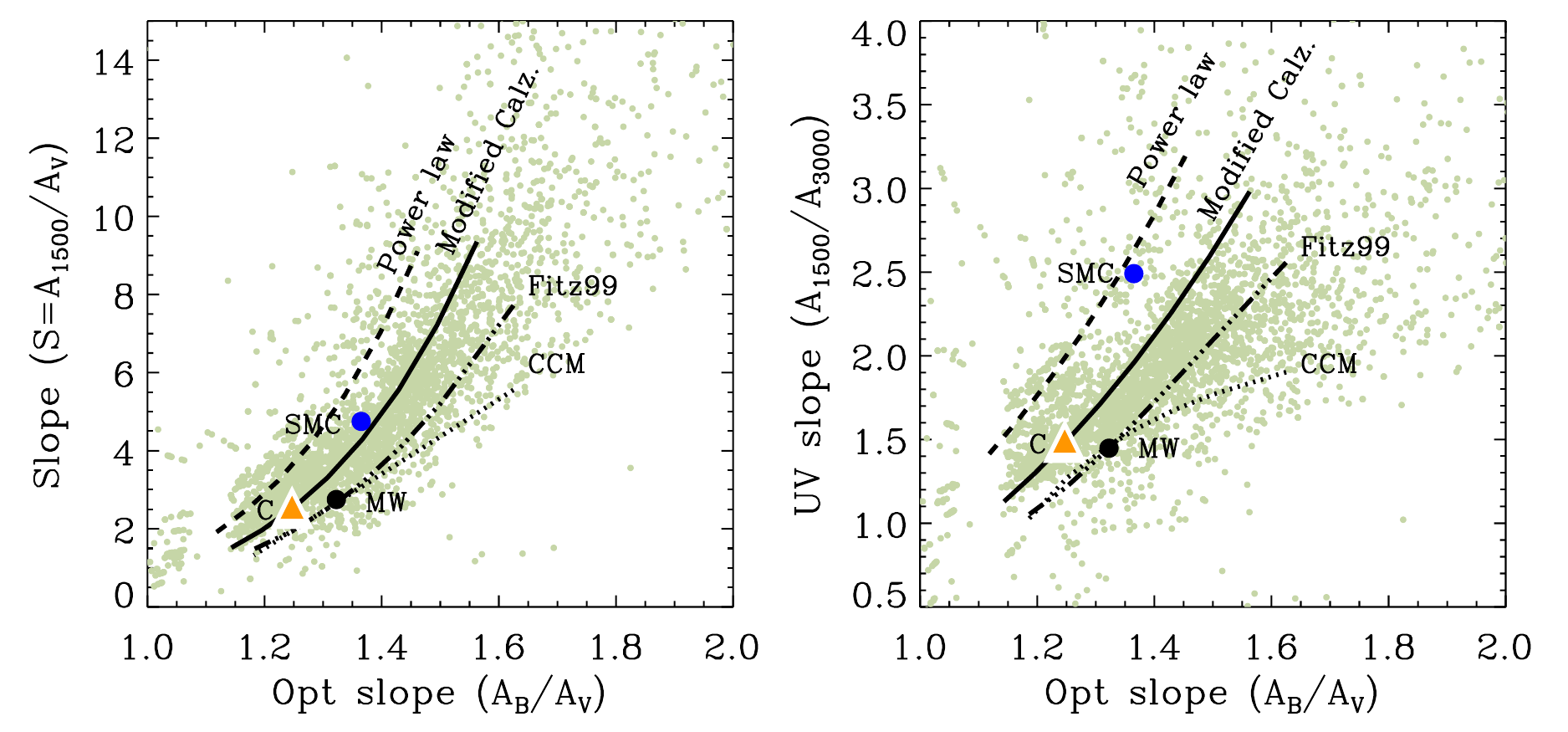}
    \caption{Tracks corresponding to four different $1$-parameter parameterizations overlaid on simulated attenuation curves from \citet{Narayanan2018b} that are based on the MW input extinction curve. Power-law curves span $0.5<n<1.7$, modified Calzetti curves span $-1.0<\delta<0.4$, and the \citet{Fitzpatrick1999} and CCM curves span $1.6<R_V<5.5$.  While it is not shown, modeling of dust attenuation using the age-selective (two-component) model, such that the tracks shown are applied to young and old populations separately,  will result in effective attenuation curves that are shifted (mostly along the tracks) and also increase the scatter. Positions occupied by MW and SMC extinction curves and the Calzetti attenuation curve (labeled `C') are shown for reference.  \label{figure:param}}
\end{figure}

Ideally, one wishes to determine an effective attenuation curve spanning a range from the Lyman continuum to the near-IR, and with sufficient spectral resolution to reveal any small-scale features. This is not possible with broad-band photometric data. Furthermore, when using the SED fitting method, it is necessary to describe the attenuation curve with a small number of parameters, which gives rise to various proposed parameterizations.

Many common parameterizations consist of a single parameter related to the overall slope of the curve, and, optionally, one or more parameters specifying the UV bump.
The first two parameterizations discussed here were originally developed to describe extinction curves in the Galaxy, but can in principle also serve as parameterizations for attenuation curves.

We start our discussion with the CCM parameterization of MW extinction curves  introduced in Section \ref{section:ext_obs}. The detailed shape of these curves is specified by four dozen parameters (coefficients), of which one, the optical slope ($R_V$), is allowed to vary. In the CCM parameterization the UV bump is not adjustable, but this can be achieved with appropriate modifications \citep{Conroy2010b}. As far as we know, the CCM parameterization is not currently implemented in any publicly available SED fitting code, but has been used in some studies (e.g., \citealt{Hagen2017}).

Another single-family parameterization originally derived for MW extinction curves is that of  \citet{Fitzpatrick1999}, which extends to the optical and near-IR a very flexible parameterization developed by \citet{Fitzpatrick1988} for spectroscopic extinction curves in the UV. In its full form this parameterization contains four parameters related to the overall shape of the curve, but they can be made internally dependent so that only $R_V$ is allowed to vary, leading to a single family of curves. In addition, there are three parameters that specify the strength, the central wavelength and the width of the UV bump, based on a Drude profile \citep{Fitzpatrick1986}.

Moving away from extinction curve parameterizations, \citet{Noll2009} developed a parameterization in which they took the \citet{Calzetti2000} attenuation curve and allowed it to have a variable slope and a Drude function profile to describe the bump. The parameter used to produce a family of curve slopes is the power-law exponent $\delta$, such that $\delta=0$ corresponds to the unmodified Calzetti curve, and $\delta<0$ produces curves that are steeper than it. The UV/optical slope of the curve $S$ and $\delta$ are related as:
\begin{equation}
    \log S = 0.40-0.55\,\delta \label{eqn:delta}
\end{equation}
The modified Calzetti curve parameterization is currently implemented in the SED fitting code {\sc cigale} in a form described in \citet{Boquien2019} and \citet{Salim2018}. It is also available in {\sc prospector} \citep{Leja2017}, but with the bump tied to the value of the slope according to a relation from \citet{Kriek2013}.

All of the previous parameterizations produce curves whose shapes are based on empirical extinction or attenuation curves. A less nuanced approach is to describe the attenuation curve as pure power law, which also requires one parameter ($n$).

In Figure \ref{figure:param} we show the four aforementioned parameterizations in terms of the relationships between UV/optical slope and the optical slope (left), and just the UV slope against the optical slope (right). For illustration purposes we also show simulated attenuation curves from \citet{Narayanan2018b}. Some parameterizations follow the simulations better than others, but with the important caveat that \citet{Narayanan2018b} assume a  single extinction curve as an input (that of the MW) and true curves may fill the parameter space differently as grain size distributions and compositions vary from galaxy to galaxy. The bottom line is that each of the four parameterizations covers a significantly different part of the parameter space and, being single-parameter curves, none can capture the diversity of the simulated curves. Some broadening of the loci ($\sim$0.07 in $A_B/A_V$), can be accomplished when the curves are applied in a two-component fashion, i.e., with different normalization for young and old stars (Section \ref{section:birthclouds}.) 

The simulations of \citet{Narayanan2018b} do not extend in the near-IR to inform us about the performance of parameterizations in that regime, but if we use the MW extinction curves from \citet{Fitzpatrick2009} as a guide, either the family of  modified Calzetti curves or the power-law curves are likely to be more realistic than the CCM or \citet{Fitzpatrick1990} parameterizations, for no other reason than because the latter are essentially fixed curves in this wavelength range.

%% file: atten_obs_lowz.tex
\begin{figure}
    \centering
    \includegraphics[scale=0.67]{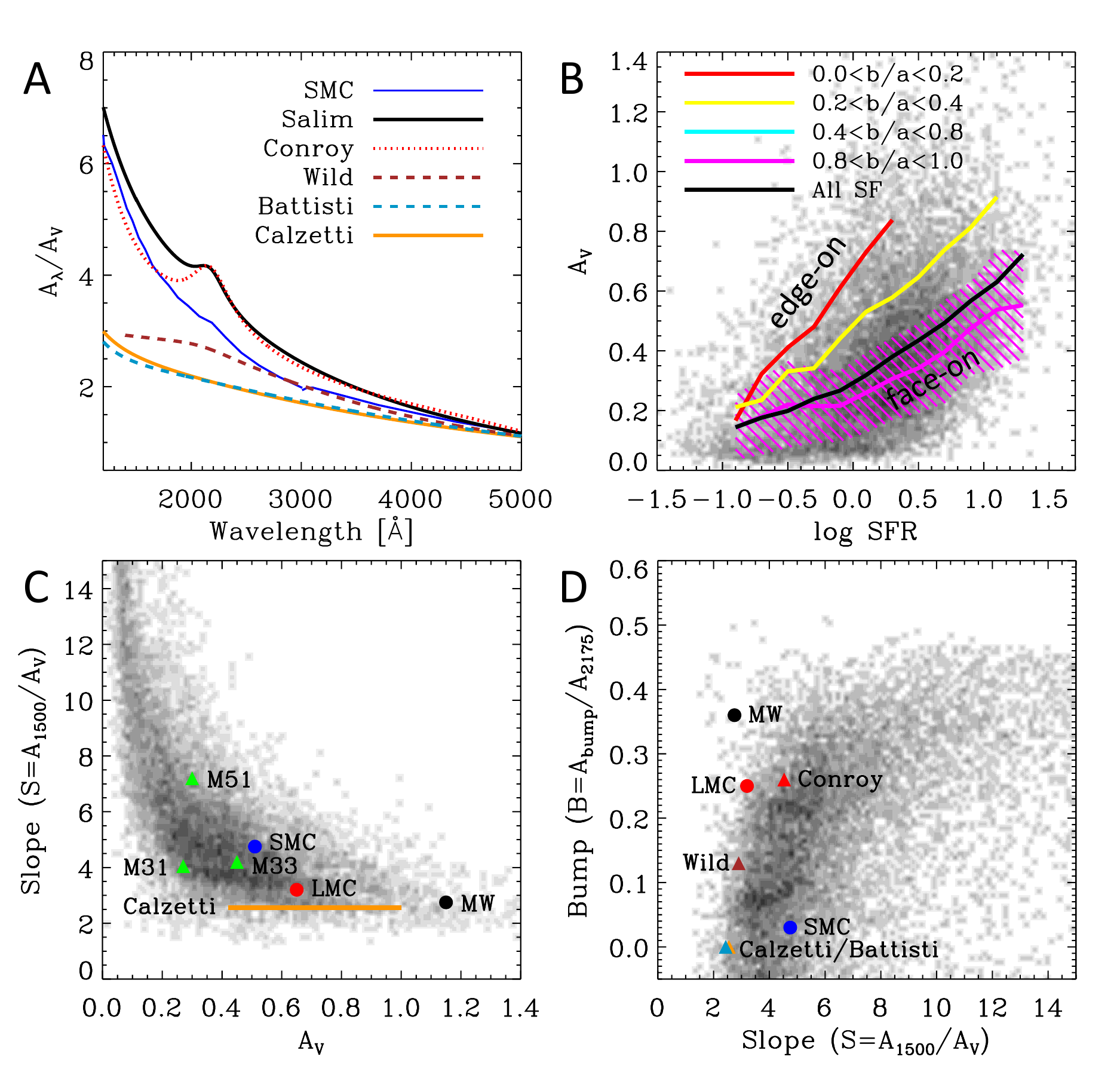}
    \caption{Observational characteristics of attenuation curves in the local universe. Panel A shows the attenuation curves of \citet{Calzetti2000,Conroy2010b,Wild2011,Battisti2016} and \citet{Salim2018} in comparison with the average SMC extinction curve. The Wild et al.\ curve is given for disk morphology, $b/a=0.6$ inclination and log sSFR$=-10$. The original Salim et al.\ curve has been rederived based on deep UV photometry from \citet{Salim2019}.
    Panel B shows how the optical depth of galaxies changes as a function of SFR and galaxy inclination.  Panel C shows the relationship between UV/optical slopes and $A_V$ for individual galaxies, and Panel D shows the UV bump strength vs.\ the slope. Data for panels B, C, and D come from the deep UV sample of Salim et al. The Milky Way, LMC, and SMC represent mean extinction curves, whereas M31, M33, and M55 are attenuation curves from detailed RT modeling. \label{figure:obs_lowz1}}
\end{figure}

\section{\MakeUppercase{Attenuation Curves: Observational Results}}
\label{section:attenuation_observations}
In previous sections we built the theoretical foundation of  extinction and attenuation laws, showed model results, and surveyed the landscape of methods for deriving attenuation curves. We now discuss the observational results relating to attenuation curves at low and high redshift.

In short, we aim to aggregate results regarding variations in the slopes ($S=A_{1500}/A_V$) and bump strengths ($B=A_{\rm bump}/A_{2175}$) in attenuation laws.  For reference, the mean Galactic extinction curve has $S_{\rm MW}=2.8$ and $B_{\rm MW}=0.36$, whereas the mean SMC extinction curve has $S_{\rm SMC}=4.8$ and $B_{\rm SMC} \approx 0$, based on the standard curves from Section \ref{section:ext_obs}.  We will refer to curves with $S>4$ as ``steep."

We subdivide results based on two factors:
\begin{enumerate}
 \item {\bf Redshift:} We split the discussion into low redshift ($z<0.5$) and higher redshift ($z>0.5$), between which there is on average an order of magnitude difference in sSFRs and gas fractions, and a significant difference in the availability and sensitivity of the data.
\item {\bf Method of deriving curves:} As we will demonstrate, different methods often produce differing results with respect to the slopes and bump strengths, even for the same sample.
\end{enumerate}

While reviewing the results we will keep in mind that the expected diversity in attenuation curves from theory (Section \ref{section:attenuation_theory}) implies that the results may differ even for same redshift and method depending on sample selection.

\subsection{Attenuation Curves of Galaxies at Redshifts Less Than Approximately 0.5}
\label{section:lowz_empirical}

We first point the reader to Figure~\ref{figure:obs_lowz1}, where we summarize the slope ($S$) and bump strength ($B$) parameters obtained from several low-redshift studies. Figure~\ref{figure:obs_lowz1}b highlights a large range of intrinsic (face-on) dust columns of galaxies ($0<A_V<0.6$) and how $A_V$ is magnified by the viewing geometry. 

\subsubsection{Empirical Methods} 

\citet{Calzetti1994,Calzetti2000} presented seminal work in deriving attenuation curves for nearby galaxies.  Indeed, the mean curve derived from these studies has been a standard reference in the $\sim$2 decades that have followed their publication, and the Calzetti curve is a common assumption in SED fitting models.
The \citet{Calzetti1994} sample consisted of 36 starbursting and blue compact dwarf galaxies with UV (from {\it IUE}) and optical spectra. The sample lies on average 1 dex above the local star-forming main sequence (the SFR-$M_*$ relation), which has motivated the use of the resulting curve at high redshift, given the increased normalization of sSFR with redshift \citep[e.g.,][]{Rodighiero2011,Speagle2014,Madau2014}.
 After anchoring the relative curve using IR luminosities (Section \ref{section:templates}), the nominal Calzetti curve is relatively shallow ($S=2.6$), similar to the slope of MW extinction curve (Figure \ref{figure:obs_lowz1}a), with a $1\sigma$ range on the slopes of $2.3<S<2.9$ \citep{Calzetti2000}.  An important feature of the Calzetti curves is the significantly reduced (or non-existent) 2175 \AA\ UV bump in the average curve. The absence of an apparent bump in the individual {\it IUE} UV spectra of 143 galaxies, which include some 20 normal star-forming galaxies and from which the Calzetti sample was drawn, was previously noted by \citet{Kinney1993}.

The subsequent introduction of large surveys, in particular SDSS in the optical and {\it GALEX} in two UV bands, allowed the empirical study of attenuation curves to be extended to larger samples and to more normal star-forming galaxies. Similarly, the 2MASS and UKIDSS surveys facilitated anchoring for curves by extending measurements to the NIR.  For example, \citet{Battisti2016,Battisti2017a} adapted the template method of Calzetti to use UV photometry, and combining this with NIR photometry obtained aggregate curves based on 5500 galaxies with 4000 \AA\ break $<1.3$ (corresponding to ${\rm sSFR}<10^{-10}$ yr$^{-1}$), i.e., relatively young main sequence galaxies. Their average curve has essentially the same shallow slope  ($S= 2.4$) as the Calzetti curve. \citet{Battisti2017a} also studied the average curves of subsamples as a function of stellar mass, sSFR and gas metallicity, but found no convincing trends, supporting the notion that the average attenuation curves of normal starforming galaxies and starbursting ones have similar slopes. Splitting the sample by inclination, \citet{Battisti2017b} did find a correlation with the attenuation curve slopes, in the sense that highly inclined galaxies have shallower slopes ($S=2.2$) with a possible weak bump ($B\sim 0.1$), whereas face on galaxies had steeper slopes (though still only $S=2.7$). \citet{Battisti2016} tentatively concluded based on broad-band UV colors and their relationship with the Balmer decrement that the UV bump was not seen in their sample. They later strengthened this conclusion based on a small sample for which {\it GALEX} UV spectra were available \citep{Battisti2017b}. 

Other empirically-based studies confirmed the relatively shallow slopes for normal star-forming galaxies.  For example, the empirical method used by \citet{Johnson2007} finds relatively shallow ($S > 2.5$)  slopes for $1000$ nearby galaxies split into several mass and 4000 \AA-break bins,
without obvious mass or age dependence. \citet{Wild2011} use an empirical pair method (Section \ref{section:templates}) to derive optical-near-IR (and UV-near-IR curves) for a sample of 23,000 (15,000) normal star-forming galaxies. They present results as a function of inclination, sSFR and central stellar mass density. They find the near-IR curve to be constant, whereas the optical and UV curves to depend on inclination and morphology, though only weakly on the sSFR and optical depth. In particular, the strength of the UV bump was found to be stronger in edge-on galaxies. Overall, the UV-optical slope of various subsamples in their analysis varies between $2.5<S<4.5$ (i.e., mostly shallow), with a moderate bump ($B\sim 0.15$).

Overall, empirical studies find a possible diversity of attenuation curves, with trends depending on inclination but not (strongly) on sSFR (i.e., population age). On average, curves derived via empirical methods tend to be shallower than the SMC extinction curve. The evidence for UV bumps is mixed, but on average they do not seem to be as strong as the MW bump.  

\subsubsection{The IRX-$\beta$ Method}
\label{section:lowz_irxb}

As demonstrated in Figure~\ref{figure:irxb_attenuation}, the tight mapping between the position of a galaxy in the IRX-$\beta$ plane and the shape of the attenuation curve allows one to deduce the properties of attenuation curves (in particular its slope; Section \ref{section:irx_method}).

The original study of starbursting galaxies with {\it IUE}  found the IRX-$\beta$ relation to be relatively narrow \citep{Meurer1999}. As UV observations with {\it GALEX} became available, it was recognized that galaxies with more normal SF had a larger scatter compared to the Meurer sample, and/or are offset from it \citep{Buat2005,Seibert2005,GildePaz2007,Dale2009,Takeuchi2010}. Several studies have subsequently rederived the Meurer relation using {\it GALEX} data \citep{Overzier2011,Takeuchi2012,Casey2014b} and found it to be shifted lower, due to a factor of $\sim$2 aperture losses in {\it IUE} UV photometry, the reason previously suggested by \citet{Cortese2006}, \citet{GildePaz2007}, and \citet{Boissier2007}. As pointed out by \citet{Seon2016}, the new relations mostly eliminated the offset between starbursting and normal galaxies, suggesting that the attenuation properties of starbursting and star-forming galaxies are actually similar (namely their curve slopes). 

A number of studies have investigated stellar age-related parameters as a source of scatter in the IRX-$\beta$ relation of normal galaxies.  \citet{Kong2004} attributed the dominant drivers of scatter to age-related parameters, such as the mean population age, sSFR, or birthrate parameter (see also \citealt{Grasha2013}). This said, many subsequent works (e.g., \citealt{Burgarella2005,Seibert2005,Johnson2007,Panuzzo2007,Boquien2009,Salim2019}) did not find the age to be an important factor in driving the IRX-$\beta$ scatter within the population of galaxies they studied. The interpretation of these results would be that the slope of the attenuation curve is not strongly dependent on the age.

Moving away from regular starbursts and normal star-forming galaxies, numerous works have demonstrated that ultraluminous infrared galaxies (ULIRGs) tend to lie much bluer than the starburst relation \citep[e.g.][]{Goldader2002,Buat2005,Howell2010,Casey2014b}, suggesting curves similar to or shallower than the Calzetti relation. Indeed, \citet{Casey2014b} found that galaxies with increasing infrared luminosity at a given redshift have increasingly blue IRX-$\beta$ colors at both low and high-redshift, a result subsequently postdicted by theoretical simulations \citep{Safarzadeh2017,Narayanan2018b}. The physical interpretation forwarded by \citet{Casey2014b,Safarzadeh2017}, and \citet{Narayanan2018b} is that low optical depth sightlines toward unobscured OB associations drive bluer colors.

\subsubsection{Model-Based Methods}

\citet{Burgarella2005} presented a seminal study deriving attenuation curves from nearby galaxies via SED fitting methods.  These authors analyzed $50$ UV and $100$ IR selected galaxies, and based on a single-population power-law parameterization that allows for a bump, found that individual galaxies span a range of slopes ($2.5<S<6.2$) and a range of bump strengths from zero to almost MW-like. On average, their UV selected sample had $S=4.5$, almost as steep as the SMC, with a moderate bump ($B=0.2$), i.e., half as strong as the MW bump.   
\citet{Conroy2010b} applied a color-based method and two-component power-law model to a principal sample of 3400 medium-mass ($9.5< \log M_*<10$) disk galaxies ($\mathrm{Sersic\ index}<2.5$) to derive an average curve with the same steep slope ($S=4.5$) as in \citet{Burgarella2005}, but a somewhat stronger bump of $B=0.26$. As part of the demonstration of the Bayesian Markov Chain Monte Carlo (MCMC) SED fitting code {\sc prospector}, \citet{Leja2017} performed UV to mid-IR SED fitting on a diverse sample of 130 relatively massive galaxies, and found a wide range of slopes in the $2<S<15$ range.  Recently, \citet{Salim2018} applied UV/optical+IR SED fitting to obtain individual attenuation curves of 230,000 SDSS galaxies, including the galaxies below the main sequence. Curves were found to exhibit a very wide range of slopes $2<S<15$, with the median $S=5.4$, somewhat steeper than the SMC curve (Figure \ref{figure:obs_lowz1}a and c).

Both \citet{Leja2017} and \citet{Salim2018} find a strong relationship between the optical opacity and curve slope, such that the galaxies with higher $A_V$ have shallower slopes. We show this relationship for individual galaxies in Figure~\ref{figure:obs_lowz1}c, using the deep-UV subset of \citet{Salim2018} sample. Also shown are the slopes for M51, M31, and M33, determined from detailed 3D radiative transfer modeling  \citep{DeLooze2014,Viaene2017m31,Williams2019}, 
which generally fall on the relation. Also shown are the slopes from standard extinction curves for the SMC, LMC, and MW. The MW point has high $A_V$ because typical sightlines are in the Galactic plane. 

The UV bump was also found to vary significantly in the \citet{Salim2018} derived curves, from no bump to values somewhat higher than the MW's bump, and with a median value of $B=0.2$ (half the MW strength). These results support the existence of a correlation between the slope and the bump strength (Figure \ref{figure:obs_lowz1}d), previously found by \citet{Kriek2013}, and in agreement with the simulations of \citet{Narayanan2018b} (Figure \ref{figure:sim}). The sense of the relation is that shallower slopes tend to have weaker bumps.  Notably, it is the opposite of the trend in the SMC, LMC, and MW extinction curves, and also differs from the \citet{Battisti2017b} derived attenuation curves, in which there are signatures of a bump only in the most inclined galaxies (those with shallower curves).

Recently, \citet{Decleir2019} has applied the SED+LIR method using UV photometry from {\it Swift} in addition to {\it GALEX} to study the attenuation curve of the low-mass spiral galaxy NGC628 (M74) on global and resolved scales. The overall curve of this $A_v=0.26$ galaxy is steep ($S=5.3$), with a moderate bump ($B=0.14$). Interestingly, individual regions also show slope vs.\ $A_V$ and slope vs.\ bump strength relations. {\it Swift} photometry was found to significantly help constrain the curve, especially the UV bump.

As a method that yields attenuation curves for individual galaxies, the model based studies have the potential of presenting us with the detailed demographics of dust properties and how they relate to other galaxy properties. Like empirical method studies and IRX-$\beta$ studies, the model-based studies agree that there is no large difference in the average slopes of starburst and normal-star-forming galaxies. Where model-based studies differ from empirical studies is that they find on average steeper slopes for similar samples of galaxies. There is tentatively also a discrepancy regarding the average strength of the bump:  model-based studies find moderate bumps (though not as strong as in the MW attenuation curve), while empirical studies find weak or no bumps. More work is needed to clarify these methodological and/or observational issues.

\begin{figure}
    \centering
    \includegraphics[scale=0.7]{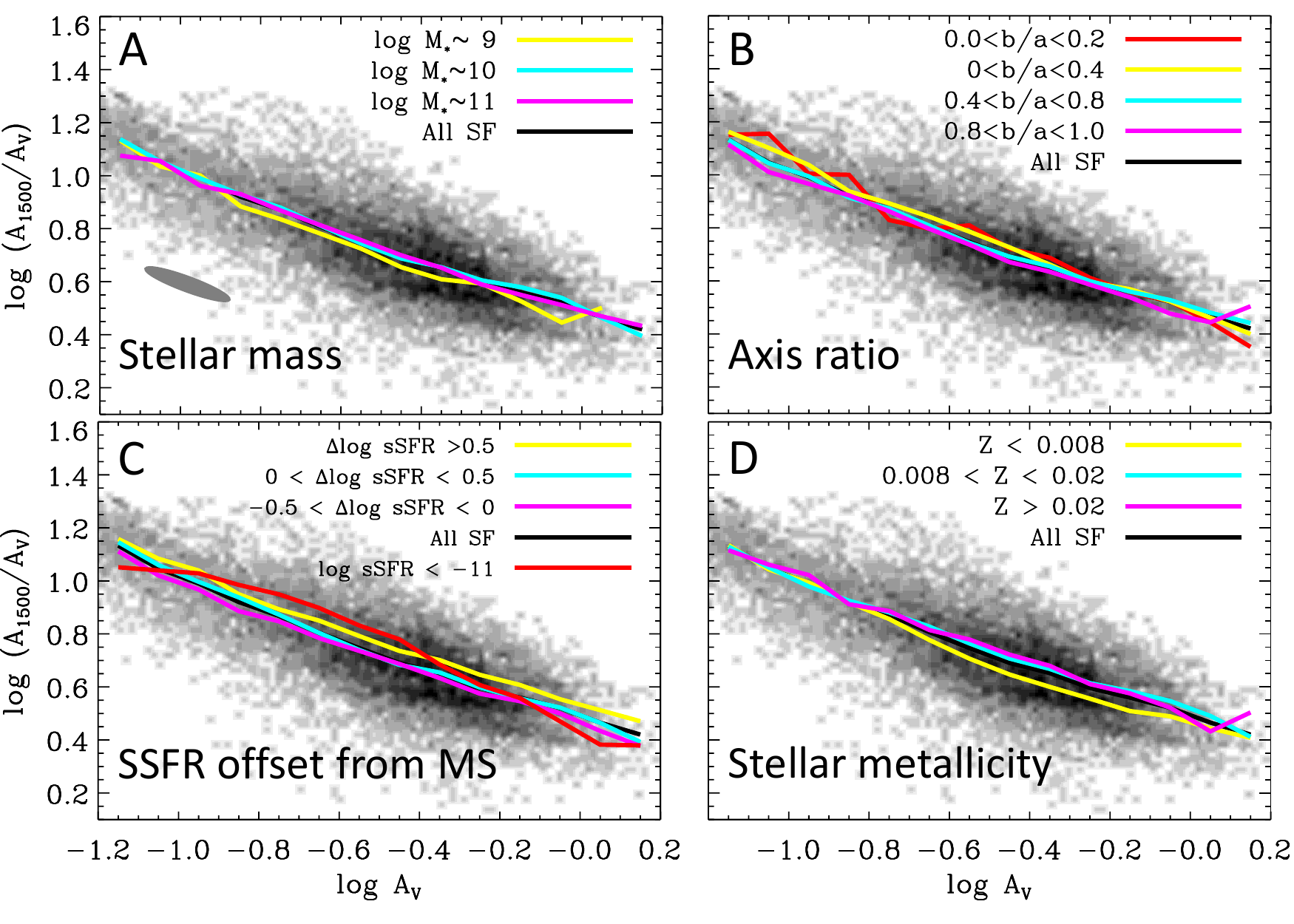}
    \caption{Dependence of the slopes of attenuation curves on galaxy physical parameters at $z\sim 0$. Log-log plots of the UV/optical slope vs.\ $A_V$ are based on the same data used in Figure \ref{figure:obs_lowz1}. Each panel contains the same data and the same average trends for the full sample (black curves). Individual panels show trends for subsets of data binned by: stellar mass, inclination (axis ratio), sSFR offset from the main sequence (positive means above) and the stellar metallicity. Except for the metallicity, none of the other factors affects the slope of the attenuation curve directly, but only by modifying $A_V$. The marginalized errors are $\sim$0.1 dex  in both coordinates, and are highly correlated (ellipse in panel A). \label{figure:obs_lowz2}}
\end{figure}

\subsubsection{Drivers of the slope of attenuation curves in the local universe}

Large samples of attenuation curves enable systematic investigations of the drivers of their diversity.  Here, we use 23,000 galaxies with deep UV from {\it GALEX}-SDSS-{\it WISE} Legacy Catalog 2 (GSWLC-2, \citealt{Salim2019}) and plot the slopes $S$ vs.\ $A_V$ in Figure~\ref{figure:obs_lowz2}.  First, we see that the slope-$A_V$ relation has the power-law form:

\begin{equation}
\log S = \log (A_{1500}/A_V) = -0.680\, \log A_V + 0.424.
\label{eqn:slope_av}
\end{equation}

\noindent The scatter around the relation is 0.12 dex, and most of it is intrinsic. This empirical relation is not as steep as the one predicted in modern cosmological simulations (cf. Figure~\ref{figure:sim}c and \citealt{Narayanan2018b}), though it is qualitatively similar. The difference may arise from a higher fraction of young stars in dense regions in real galaxies \citep{Inoue2005}. The principal question can be stated simply: what parameters other than the dust column density (for which $A_V$ serves as a proxy here) may affect the shape of the attenuation curve, in particular its slope? The panels of Figure \ref{figure:obs_lowz2} show trend lines for subsamples of star-forming galaxies of differing (a) stellar mass, (b) inclination, (c) specific SFR with respect to the main sequence, and (d) stellar metallicity. The key point is that once $A_V$ is fixed, none of the investigated factors has a significant effect on the slope of the attenuation curve. For example, more massive galaxies on average do have shallower curves, but this is because more massive galaxies have higher $A_V$, i.e., they are shifted down the relation. Remarkably, the lack of difference in the slopes at fixed $A_V$ is true even for galaxies with very different inclinations. A dusty galaxy with a face-on attenuation of $A_V$ and a less dusty galaxy that has the same $A_V$ (because we observe it edge-on) will have the same attenuation curve slope, as predicted in the radiative transfer models analyzed by \citet{Chevallard2013}. A similar situation exists for different levels of star formation activity: galaxies from starbursts to green valley and quiescent galaxies below the main sequence (red line in Figure \ref{figure:obs_lowz2}c) all have very similar slopes for a given $A_V$. A small trend may exist with respect to the metallicity (both stellar and gas-phase), such that more metal-rich galaxies have steeper slopes at fixed $A_V$, which may be a reflection of different dust grain properties, rather than radiative transfer effect. Galaxies with different bump strengths (not shown) will have different slopes at a fixed $A_V$, in the sense that weaker bumps go with shallower curves. However, rather than being the driver of the dispersion, the trend reflects the fact that low $A_{1500}$ helps suppress the bump, as discussed in Section \ref{section:theory_bump}.

Residual intrinsic scatter of slopes at fixed $A_V$ may be due to the differences in the intrinsic dust grain properties (i.e., the size distribution or composition), which is difficult to test. However, if the slopes of the extinction curves themselves follow $A_V$, as suggested in Sections \ref{section:extinction_parameters} and \ref{section:extinction_extragalactic}, such that low $A_V$ sightlines suffer steeper (SMC-like) extinction, we can speculate that the diversity of ``dust types" at a given $A_V$ may potentially be small. 

We conclude that the dominant determinant in the slope of the attenuation curve, out of the considered physical parameters, is $A_V$. Galaxies with different inclinations, star formation rates, and stellar masses to first order ``move" up and down the more or less universal relation, depending on their $A_V$ values.

%% file: atten_obs_highz.tex
\subsection{Attenuation Curves of Galaxies at Redshifts Greater Than Approximately 0.5} 
\label{section:atten_highz}

\begin{figure}
    \centering
    \includegraphics[scale=0.65]{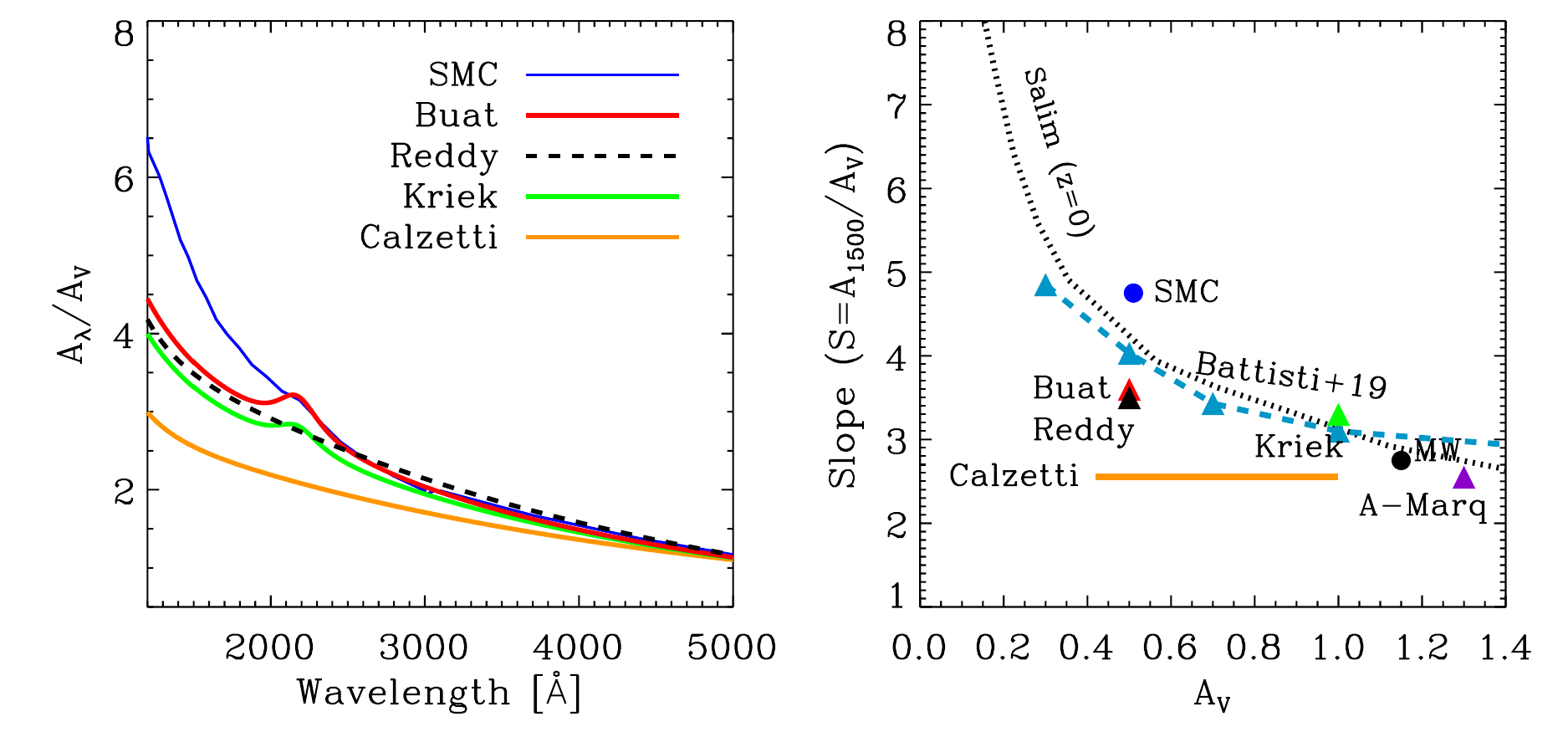}
    \caption{Observational characteristics of attenuation curves at $z>0.5$.  {\bf Left:} Attenuation curves at $z>0.5$ from  \citet{Buat2012,Kriek2013}, and \citet{Reddy2015} in comparison to the SMC and Calzetti curves. The \citet{Buat2012} and \citet{Kriek2013} curves  were constructed using the SED fitting method, whereas \citet{Reddy2015} use the empirical templates-based method. {\bf Right:} The slope of curves compared with sample-averaged $A_{\rm V}$. This panel adds data points for galaxies split by $A_V$ from \citet{Battisti2019} and the data point from \citet{Alvarez2019}. \label{figure:obs_highz}}
\end{figure}

Galaxies at high-redshift differ in multiple ways from the present-epoch galaxies: at a fixed stellar mass they have larger gas fractions and star formation rates, their morphologies and kinematics are less settled, and their gas-phase metallicities are lower (see the reviews by \citealt{Shapley2011} and \citealt{Madau2014}). All of these factors may affect the dust content, properties of dust grains, and the resulting dust attenuation curves.  The methodological and observational difficulties that affect the determination of average or individual attenuation curves at low redshift are exacerbated at higher redshifts due to smaller sample sizes, less sensitive or less complete photometry (especially in the infrared), the reliance on photometric redshifts, more difficult identification of AGN contamination, and other related issues. Informed by low-redshift studies, we will pay special attention to the properties of the samples being studied in order to account for the differences that may stem from the diversity of high-redshift populations. As we will show, the general picture that is starting to emerge from studies of galaxies at high-redshift is one that is in broad terms consistent with what is observed at low-redshift: galaxies exhibit a wide range of attenuation curve slopes, from grey Calzetti-like curves to steeper than SMC curves.  

Several high-redshift studies have found a correlation between the curve slopes and optical attenuation, similar to the results at low redshift.  Potentially as a result of this relation, the IR-selected galaxies tend to exhibit shallower overall curves than UV or optically-selected galaxies, with optically nearly opaque star forming galaxies having the grayest curves. The UV bump is present at a variety of strengths, and is on average correlated with the slope. Figure \ref{figure:obs_highz} summarizes the results of a number of high-redshift studies.

\subsubsection{Nebular vs.\ stellar continuum attenuation}
Before embarking on the review of results related to attenuation curves, we highlight an important topic that has emerged in high-redshift studies: the quantification of the relative attenuation toward ionized regions vs.\ the stellar continuum (reviewed by \citealt{Steidel2014} and \citealt{Shivaei2015}).

As found in the original \citet{Calzetti1994} study, evidence for increased attenuation toward nebular regions may indicate the presence of a birthcloud-like dust component \citep[e.g.,][]{Charlot2000}. These results were confirmed for low-redshift galaxies in a large-scale study of \citet{Wild2011}.
The results for the need for extra attenuation toward H{\sc ii} regions at high-$z$ are mixed. \citet{Reddy2004, Erb2006,Reddy2012}, and \citet{Shivaei2015} find that the reddening toward nebular regions is comparable to that of the stellar continuum when comparing $X$-ray, UV, H$\alpha$, infrared, and radio-based star formation rate indicators.   At the same time, \citet{Forsterschreiber2009,Wuyts2011,Price2014}, and \citet{Mieda2016}  find the need for extra attenuation toward nebular regions, consistent with the two-component birthcloud model.  \citet{Forsterschreiber2009} compared against some of the same galaxies as \citet{Erb2006}, and suggested that aperture corrections may owe to the differing results regarding potential excess attenuation toward H{\sc ii} regions.  An important step forward was made by \citet{Price2014}, who found not only the need for increased attenuation toward H{\sc ii} regions in their sample of $\sim$150 $z\sim 1.5$ galaxies drawn from the 3D-HST survey \citep{Brammer2012}, but also a decreased difference in $A_\text{H\sc{ii}}$ compared to $A_{\rm UV}$ with increasing sSFR (also \citealt{Puglisi2016}).  The physical interpretation is that for galaxies with very large sSFR, most of the stellar light is coming from young stars that see both dust in H{\sc ii} regions, as well as diffuse dust.  In these situations, $A_\text{H\sc{ii}} \approx A_{\rm UV}$.  With decreasing sSFR, however, the contribution of older stellar populations to the integrated light outside of H{\sc ii} regions becomes important, and the relative difference in reddening between H{\sc ii} regions and the stellar continuum increases accordingly. A trend with sSFR indeed exists even at low redshift \citep{Wild2011}, with the ratio of $A_B-A_V$ reddenings for the nebular and the continuum reaching equality for extreme local starbursts ($\log \text{sSFR} \sim -8.7$, \citealt{Koyama2019}).

Finally, we point out that the comparison of nebular and stellar continuum attenuation requires that the attenuation curves are  known or correctly assumed for both components. \citet{Wild2011a} have found that the nebular attenuation curve has a slope similar to the MW extinction curve, rather than the SMC or the Calzetti curves, so the standard Galactic extinction law is recommended for correcting emission line fluxes.

\subsubsection{Empirical methods} 

Studies using empirical methods (Section \ref{section:templates}) have placed a number of constraints on the slopes of high-$z$ attenuation curves.   Early work by \citet{Reddy2004} found consistent results between $X$-ray and radio-derived SFRs from UV-selected galaxies at $1.5<z<3$, and those derived from dust-corrected UV photometry assuming a \citet{Calzetti1994} curve, implying a Calzetti-like curve is a reasonable descriptor of the attenuation in these galaxies.  This result was confirmed by \citet{Shivaei2015}, who used a similar method for a sample of $\sim$250 $z\sim2$ UV-selected galaxies.    

At the same time, a number of studies have found evidence for curves that depart from Calzetti-like at high redshift.  For example, using the spectroscopic data from the MOSFIRE Deep Evolution Field (MOSDEF) rest-frame optical survey of 224 $H$-band selected $z \sim 2$ galaxies, \citet{Reddy2015} applied the Balmer-decrement based template method of \citet{Calzetti1994} to derive relative attenuation curves.  These curves were anchored to absolute curves in several ways. Assuming zero attenuation at 2.9 $\mu$m resulted in $R_V=2.51$, which was adopted as fiducial curve, and we show it in Figure \ref{figure:obs_highz}. With the slope of $S=3.5$, the curve sits roughly halfway in between the Calzetti ($S=2.6$) and the SMC curves ($S_{\rm SMC}=4.8$), for a sample that has an average $A_{\rm V} = 0.5$.

\citet{Reddy2015} find marginal evidence for the UV bump using broadband data, though the authors do not include it in their third-order polynomial attenuation curve fit. \citet{Noll2007} use a sample of 108 UV-selected galaxies also at $z\sim 2$, assembled from several spectroscopic surveys. By looking at the difference in SED slopes shortward and longward of the bump peak, they find that galaxies on average have bump strengths close to that of the LMC 30 Dor extinction curve ($B=0.15$) and essentially never as strong as the MW bump ($B_{\rm MW}=0.36$). Similar results were obtained by \citet{Noll2009b} in a follow-up study of 78 galaxies in the same redshift range. In addition, they found that the width of the composite bump was 40\% narrower than in the MW. \citet{Conroy2010c} found that the broad-band colors of $z \sim 1$ galaxies are inconsistent with bumps as strong as the Galactic one if a shallow curve with MW-like slope is assumed.  This said, a moderate bump in the \citet{Conroy2010c} is not excluded if the slope is steeper.

\subsubsection{The IRX-$\beta$ method} 
\label{section:irxb_highz}

The principal challenge with the IRX-$\beta$ method at higher redshifts lies in the fact that typical galaxy populations are too faint to be individually detected, especially at longer IR wavelengths. More recently, this limitation has been addressed by the application of stacking techniques and observations using ALMA, and we will primarily focus on those studies. Like low-redshift galaxies, galaxies at higher redshifts occupy a wide range of positions on the IRX-$\beta$ diagram. Many studies find that high-redshift galaxies on average follow the original Meurer relation uncorrected for UV aperture losses \citep[e.g.,][]{Siana2009,Bourne2017,Fudamoto2017,McLure2018,Koprowski2018,Wang2018}. The interpretation of these results would be that, on average, these galaxies have attenuation curves similar to the Calzetti curve.  Other studies find IRX-$\beta$ relations that, when interpreted as attenuation curves, place them closer to the SMC curve \citep[e.g.][]{Heinis2013,Alvarez2016,Pope2017,Reddy2018}. \citet{Reddy2018} further find that lower-mass galaxies tend to have somewhat steeper attenuation curve slopes. The cause of the differing results is as yet not clear, but could in principle arise from the differences in typical dust contents of the samples that have been selected in different ways. Furthermore, as pointed out in Section \ref{section:irx_method}, the interpretation and comparison of the IRX-$\beta$ results is made difficult by different methods of measuring the UV slope (e.g., \citealt{Koprowski2018,Alvarez2019}) and the uncertainties in the magnitude of evolution of the intrinsic UV slope \citep{Buat2012,Reddy2018}. 

In addition, there are notable deviations in how galaxies populate the IRX-$\beta$ plane that go beyond the Calzetti-to-SMC range. They come in two general flavors: a cloud of galaxies that are particularly blue (i.e., low $\beta$), which lie above and to the left of the IRX-$\beta$ relation corresponding to the Calzetti curve, and galaxies that have IRX values which place them below the SMC curve.  Galaxies with blue UV spectral slopes are often selected as dusty star forming galaxies \citep[DSFGs; for a review, see][]{Casey2014a}, via a range of techniques \citep[e.g.,][]{Penner2012,Oteo2013,Forrest2016}. \citet{Casey2014b} demonstrated that galaxies with increasing infrared luminosity have increasingly blue UV SEDs
and speculated that this may be due to to low-opacity sightlines toward UV-bright sources.  This physical scenario corresponds to grayer attenuation curves driving galaxies upward in IRX-$\beta$ space, as demonstrated explicitly in Figure~\ref{figure:irxb_attenuation}. 

An analogous situation going in the opposite direction may apply to galaxies detected at $z > 5$ that all appear to fall below the IRX-$\beta$ curve associated with the average SMC attenuation curve \citep{Capak2015, Bouwens2016, Barisic2017}.  While at the presence of these high-redshift systems at very low IRX values may at least partially owe to an underestimate of the infrared luminosity \citep[e.g.,][]{Faisst2017,Narayanan2018a,Liang2019,Ma2019}, some theoretical studies have found that the IRX-$\beta$ relation even for these low-metallicity systems is shaped predominantly by the attenuation curve itself \citep{Narayanan2018a,Reddy2018,Ma2019} and would therefore indicate very steep curves.

Finally, we note there has been recent interest in the IRX-$M_*$ relation at high redshift \citep[e.g.,][]{Alvarez2016,Bouwens2016,Fudamoto2017}. Locally, this relation has a scatter almost two times larger than the IRX-$\beta$ relation \citep{Salim2019}, but when the stellar mass is added to the IRX-$\beta$ relation as a $3$-dimensional plane showing the variation of IRX, $\beta$, and $M_*$ for galaxies at a fixed redshift, it can help obtain more accurate predictions for IRX, and reveal variations in the attenuation law across the stellar mass function at a given cosmic epoch. 

\subsubsection{Model-based methods} The application of SED fitting methods to derive attenuation curve parameters of high-redshift galaxies is a relatively recent method, but has revealed important correlations in the attenuation curves of galaxies.  Similarly, these techniques have elucidated the importance of the near-IR portion of the attenuation curve in deriving physical properties of galaxies \citep{LoFaro2017}.

\citet{Buat2011b} performed UV through far-IR SED fitting of 30 $z\sim1.5$ galaxies selected at 160 $\mu$m with a \citet{Noll2009} single-component parameterization. The average curve they found for these galaxies was shallow ($S=3.0$) and exhibited a moderate UV bump ($B=0.17$, 1/2 of $B_{\rm MW}$), without a correlation between the slope and the bump. This said, their IR selection resulted in a sample with a high average optical attenuation ($A_V=0.9$). Using similar techniques (with two-component dust modeling instead of one) and focusing on the same redshift range, but using a sample of 750 UV-selected galaxies of which about half are detected at 24 $\mu$m, \citet{Buat2012} find a much greater diversity of attenuation curve slopes ($2.7<S<6.3$). The geometric average is $S=3.6$, which is very similar to the \citet{Reddy2015} curve (see Figure \ref{figure:obs_highz}), with which it shares the average $A_V$ of 0.5. In the \citet{Buat2012} study, the UV bumps take a wide range of strengths, rarely exceeding that of the MW, with an average of $B=0.20$. \citet{Buat2012} found no clear trend between the slopes of their attenuation curves and the bump strengths. 

At the same time, \citet{Kriek2013} fit composite SEDs spanning from observed-frame optical to 8 $\mu$m  with the single-component \citet{Noll2009} parameterization, constructed from galaxies at $0.5<z<2.0$, and selected to have $A_V>0.5$. Their binned curves span a range of slopes from Calzetti to SMC, with an overall average of $S=3.3$. The average bump strength was moderate, with $B=0.15$ (Figure \ref{figure:obs_highz}).   An important result from the \citet{Kriek2013} study was a trend between the bump strength and the slope of composite curves, in the sense that the shallow (Calzetti-like) curves have smaller (but still detectable) bumps. In the same vein, \citet{Alvarez2019} performed SED fitting on stacks of $z\sim 3$ galaxies binned by different properties and found slopes, on average, similar to the Calzetti one.
\citet{Tress2018} performed SED fitting in the observed rest-frame optical to 3.6 $\mu$m range of 1800 galaxies at  $1.5<z<3$, exploiting 25 medium-band filters well placed to sample the UV bump. \citet{Tress2018}  found a strong correlation between bump strength and reddening, stronger than with respect to the optical slope. The strength of UV bumps was found to occupy a similar range as in previous studies discussed in this section. 
 
Most SED fitting studies find average slopes steeper than the Calzetti curve. In contrast, \citet{Scoville2015} used an alternative model-based method tailored for galaxies dominated by young stars, and found a Calzetti-like slope. Their sample lies at $2<z<6$, with an average $A_V=0.6$. The average curve derived by \citet{Scoville2015}  nevertheless shows some presence of a 2175 \AA\ UV bump. Using a similar method, \citet{Zeiman2015} obtained, for a sample of mostly lower-mass ($\log M_*<10$) galaxies at $z\sim 2$, an aggregate curve agreeing with that of \citet{Scoville2015}, except for a lack of the UV bump. 
 
As discussed in Section \ref{section:attenuation_theory}, one expects a strong  correlation between slope and $A_V$. Several studies have found this relation at higher redshifts, using attenuation curves derived from SED fitting techniques. For example, \citet{Arnouts2013} performed SED fitting with a two-component power-law dust model to 16,500 24 $\mu$m selected galaxies with $z<1.3$. They found a very wide range of slopes ($2.5<B<7.0$) with the typical value of $S=4.2$, and uncovered a correlation between the slope and the average optical opacity.  Similarly, \citet{Salmon2016} performed SED fitting (without IR constraints) on $\sim$1000 individual galaxies at $1.5<z<3$ detected at 24 $\mu$m. They too showed a diversity of attenuation curves, with similar average slopes as in \citet{Kriek2013}, and showed their slopes to be correlated with the reddening ($A_B-A_V$). Recently, \citet{Battisti2019} performed energy-balance SED fitting of $\sim$5000 IR-detected galaxies (24 $\mu$m or longer)  from GAMA and COSMOS surveys spanning $0.1<z<3$. They present attenuation curves binned by $A_V$ (excluding $A_V<0.2$), which also show a correlation with the steepness of the slope. Their results, expressed as UV/optical slope $S$, are displayed in the right panel of Figure \ref{figure:obs_highz}. Comparison with the low-redshift relation (Equation \ref{eqn:slope_av}), based on \citet{Salim2018}, suggests that the $S-A_V$ relation may not evolve strongly with redshift. The UV bumps in \citet{Battisti2019}, which are well constrained using medium-band data, are on average of moderate strength ($B=0.1$), and not correlated with other galaxy parameters.

As in low-redshift studies, differing results at higher redshifts (especially from IRX-$\beta$ studies) are present, but this may be alleviated with the consideration of the diversity of the samples, especially in terms of their dust content. Results may also depend on the methodology, and future studies should aim to resolve this question by applying several methods on the same sample and critically investigating these applications.